\documentclass[twocolumn]{jpsj2}
\usepackage{graphicx}

\def\runauthor{Takuji {\sc Nomura}}

\title{
Theory of Transport Properties in the $\mib{p}$-wave 
Superconducting State of Sr$_2$RuO$_4$\\
- A Microscopic Determination of the Gap Structure -}

\author{Takuji {\sc Nomura}\footnote{E-mail: nomurat@spring8.or.jp}}

\inst{
Synchrotron Radiation Research Center,
Japan Atomic Energy Research Institute, 
Mikazuki, Sayo, Hyogo 679-5148}

\recdate
{
}

\abst{
We provide a detailed quantitative analysis of transport properties 
in the $p$-wave superconducting state of Sr$_2$RuO$_4$. 
Specifically, we calculate ultrasound attenuation rate and 
electronic thermal conductivity within the mean field approximation. 
The impurity scattering of the quasi-particles are treated 
within the self-consistent $T$-matrix approximation, 
and assumed to be in the unitarity limit. 
The momentum dependence of the gap function 
is determined by solving the Eliashberg equation 
for a three-band Hubbard model with the realistic 
electronic structure of Sr$_2$RuO$_4$. 
On the basis of the microscopic theory, 
we can naturally expect nodal structures along the $c$-axis 
on the cylindrical Fermi surfaces, even if we assume 
the chiral pairing state (i.e., $\Delta(k) \sim k_x \pm {\rm i} k_y $). 
Consequently, we obtain the temperature dependence 
of the transport coefficients in agreement with the experimental results. 
We can clarify that actually the thermal excitations 
on the passively superconducting bands contribute significantly 
to the thermal conductivity in a wide temperature range, 
in contrast to the case of other physical quantities. 
}

\kword{
$p$-wave superconductor, Sr$_2$RuO$_4$, 
superconducting gap structure, transport properties, 
ultrasound attenuation rate, thermal conductivity
}

\begin{document}
\sloppy
\maketitle

\section{Introduction}
\label{Sc:Introduction}

The quasi-two-dimensional ruthenate Sr$_2$RuO$_4$ 
has attracted much interest of solid state physicists 
since the discovery of superconductivity 
in this compound~\cite{ref:Mackenzie2003}. 
A number of excellent experiments and theoretical considerations 
have already revealed its remarkable and intriguing physical properties. 
The most important is that Sr$_2$RuO$_4$ is a strong candidate of 
odd-parity spin-triplet superconductor 
(most likely $p$-wave superconductor)~\cite{ref:Mackenzie2003, 
ref:Ishida1998, ref:Duffy2000}, 
although the crystal structure 
of Sr$_2$RuO$_4$ is the quasi-two-dimensional 
perovskite structure identical to that of high-$T_{\rm c}$ 
copper oxides (spin-singlet $d_{x^2-y^2}$-wave superconductors). 

Mechanism of the spin-triplet pairing in Sr$_2$RuO$_4$ 
has been discussed from various microscopic theoretical points of view. 
In the early stage of the researches, it was considered 
that some strong ferromagnetic spin fluctuation (Paramagnon) 
will exist and induce the triplet pairing~\cite{ref:Rice1995,ref:Mazin1997}. 
However, the inelastic neutron scattering measurement 
showed that the spin correlation is predominantly incommensurate 
antiferromagnetic rather than ferromagnetic~\cite{ref:Sidis1999}. 
Interestingly, some recent theoretical works have shown 
that the pair-scattering amplitude could actually possess 
such a characteristic momentum dependence as 
is favorable for $p$-wave pairing but not attributable 
to that of the spin susceptibility~\cite{ref:Nomura2000b,
ref:Nomura2002b,ref:YanaseRev2003}. 
These theoretical discussions suggest that the triplet pairing 
in Sr$_2$RuO$_4$ is considered as a natural result from electron correlation, 
but the essential pairing attraction between the Fermi liquid quasi-particles 
does not originate from the exchange processes 
of spin fluctuations~\cite{ref:Nomura2000b,ref:Nomura2002b,
ref:YanaseRev2003}. 

Another interesting problem, which is attacked in the present work, 
is the detailed clarification of the superconducting gap structure. 
Superconducting gap structure is closely related to the 
temperature dependence of physical quantities in general, 
since the thermal excitations only near the gap minima 
are essential in responding to various external perturbations. 
On the theoretical side, the gap structure is a reflection 
of the momentum dependence of pairing interaction. 
Therefore, the detailed analyses of the temperature 
dependence of physical quantities could be a valuable 
test for the validity of pairing scenario. 

The gap structure of Sr$_2$RuO$_4$ is still controversial. 
The most probable pairing symmetry 
$\Delta(k) \sim (k_x \pm {\rm i} k_y ) \hat{z}$ 
was proposed by Rice and Sigrist very early 
after the discovery of the superconductivity~\cite{ref:Rice1995}. 
The muon spin relaxation experiment~\cite{ref:Luke1998} suggests 
strongly that this chiral pairing state, 
in which the time reversal symmetry is broken, 
is the most plausible pairing state 
among several candidates. 
This pairing symmetry usually results in 
isotropic gap (i.e., the absence of nodal structure). 
However, most of existing experimental data 
exhibit power-law temperature dependence~\cite{ref:NishiZaki2000,
ref:Ishida2000,ref:Bonalde2000,ref:Lupien2001,
ref:Izawa2001,ref:Tanatar2001b}, 
and indicate a nodal gap structure 
(most probably line nodes) on the Fermi surface. 
Much effort has been devoted to reconciling these 
contradictions~\cite{ref:Agterberg1997,ref:Miyake1999,
ref:Hasegawa2000,ref:Won2000,ref:Dahm2000,
ref:Zhitomirsky2001,ref:Nomura2002a}. 

In the present article, we analyze microscopically 
the temperature dependence of two transport coefficients, 
ultrasound attenuation rate and thermal conductivity, 
in the superconducting state of Sr$_2$RuO$_4$. 
The sound attenuation rate is expected to provide a strong constraint 
on the theoretical proposals of superconducting gap 
structure~\cite{ref:Moreno1996}. 
We derive the momentum and band dependence 
of the gap function by solving Eliashberg equation 
for a realistic tight-binding electronic 
structure~\cite{ref:Nomura2002a}. 
Starting with a three-band Hubbard model, 
the effective pairing interaction 
is evaluated perturbatively to third order 
in the on-site Coulomb integrals~\cite{ref:Nomura2002b}. 
In our previous works, we showed naturally 
by this theoretical framework that 
(i) the momentum dependence of the pairing interaction 
is favorable for anisotropic $p$-wave pairing, 
and $p$-wave state is obtained 
as the most probable pairing 
state~\cite{ref:Nomura2000b, ref:Nomura2002b}, 
(ii) the superconducting gap possesses 
line-node-like structures along the $c$-axis~\cite{ref:Nomura2002a,
ref:YanaseRev2003}, and (iii) the calculated specific heat 
as a function of temperature 
fits experimental data well~\cite{ref:Nomura2002a}. 

The transport coefficients of Sr$_2$RuO$_4$ 
have been calculated by other authors~\cite{ref:Graf2000,
ref:Wu2001,ref:Kusunose2002,ref:Tanaka2003,
ref:Udagawa2004,ref:Contreras2004}. 
Most of them are based on simplified isotropic Fermi surfaces 
(two- or three-dimensional), or simple gap functions 
described by one or a few harmonic functions. 
However, as discussed in the present article, 
such simplified electronic structures and gap structures 
are generally insufficient to discuss 
the experimental data of transport properties. 

For the calculation of the transport coefficients, 
we take the way of analysis which has been developed 
for some uranium compound 
superconductors~\cite{ref:Hirschfeld1986,
ref:Schmitt-Rink1986,ref:Hirschfeld1988}: 
the non-magnetic impurity scattering is treated 
by the self-consistent $T$-matrix approximation. 
The scattering is assumed to be in the unitarity limit. 
We ignore the vertex corrections. 
In contrast to the analyses for uranium compounds 
(where they considered simplified electronic structures, 
e.g., isotropic spherical Fermi surfaces), 
we take more realistic tight-binding electronic structure 
and gap structure for Sr$_2$RuO$_4$. 

For the calculation of ultrasound attenuation rate, 
we adopt the same way as Walker and collaborators 
did~\cite{ref:Walker2001}. 
In the work by Walker {\it et al.}, they succeeded in explaining 
the large in-plane anisotropy of the attenuation rate, 
and clarified that the anisotropy 
of the electron-phonon interaction is essential. 
This successful point is taken into account also in our present work. 
However, it should be noted that their gap structure 
is quite different from ours. 
They insisted that there should be point nodes 
on the Fermi surfaces~\cite{ref:Contreras2004}. 
Although we discuss based on the purely two-dimensional model, 
we consider that it is almost impossible to obtain point nodes 
or horizontal (i.e., parallel to the basal plane) line nodes. 
This is because the pair scattering amplitude can acquire 
only negligible momentum dependence along the $c$-axis, 
due to the strong two-dimensionality. 

Consequently, we obtain a consistency 
with the experimental data of the transport coefficients 
in the overall temperature range. 
The electron-phonon coupling constants are essential parameters 
for the good fitting to the experimental data. 
In addition, the anisotropy due to the lattice structure 
is also an essential ingredient for reproducing 
the thermal conductivity and the anisotropy of sound attenuation. 
Therefore the naive conjectures from the isotropic models 
and simplified gap functions are not reliable in general. 

The present article is constructed as follows. 
In \S~\ref{Sc:Theory}, the theoretical formulation is given.
We give the analytic expressions 
of ultrasound attenuation rate 
and thermal conductivity. 
In \S~\ref{Sc:Numerical}, 
the numerical result of calculated gap structure 
is provided, and then the numerical results 
of the ultrasound attenuation 
rate and thermal conductivity are compared 
with the experimental data. 
In \S~\ref{Sc:Discussion}, 
some remarks on our results are given. 
In addition, relations of the present work 
with other works are discussed. 
Finally, in \S~\ref{Sc:Conclusion}, 
we will give concluding remarks. 

\section{Theoretical Formulation}
\label{Sc:Theory}

\subsection{Three-band Hubbard model and Eliashberg equation}
\label{Sc:ModelEliashbergEq}

The Fermi surface and the electronic structure 
of Sr$_2$RuO$_4$ near the Fermi level 
are well reproduced by tight-binding fitting~\cite{ref:Oguchi1995}. 
Since the electronic density of states near the Fermi level 
is dominated by the partial density of states 
of Ru4d$\varepsilon$ orbitals~\cite{ref:Oguchi1995, ref:Singh1995}, 
the hopping parameters are considered to describe 
the transfers between the Ru4d$\varepsilon$-like Wannier orbitals. 
We take the following non-interacting Hamiltonian
(these Wannier orbitals are characterized by $\ell=\{xy,\, yz,\, xz \}$): 
\begin{eqnarray}
H_0 &=& \sum_{\mib{k},\ell,\sigma} \xi_{\ell}(\mib{k}) 
c_{\mib{k}\ell\sigma}^{\dag} c_{\mib{k}\ell\sigma} \nonumber\\
&& + \sum_{\mib{k},\sigma} \lambda(\mib{k}) 
(c_{\mib{k}\,yz\,\sigma}^{\dag} c_{\mib{k}\,xz\,\sigma} 
+ c_{\mib{k}\,xz\,\sigma}^{\dag} c_{\mib{k}\,yz\,\sigma}), 
\end{eqnarray}
where $c_{\mib{k}\ell\sigma}$[$c_{\mib{k}\ell\sigma}^{\dag}$] 
is the electron annihilation[creation] operator 
(the pseudo-momentum, orbital and spin states are denoted 
by $\mib{k}$, $\ell$ and $\sigma$, respectively), 
and the energy dispersions are 
\begin{eqnarray}
\xi_{xy}(\mib{k})&=& 2 t_1 (\cos k_x + \cos k_y) \nonumber\\
 &&+ 4 t_2 \cos k_x \cos k_y - \mu_{xy}, \\
\xi_{yz}(\mib{k})&=& 2 t_3 \cos k_y + 2 t_4 \cos k_x - \mu_{yz}, \\
\xi_{xz}(\mib{k})&=& 2 t_3 \cos k_x + 2 t_4 \cos k_y - \mu_{xz}, \\
\lambda(\mib{k}) &=& 4 t_5 \sin k_x \sin k_y. 
\end{eqnarray}
In the present work, we take the parameter set 
$t_1=-1.00$, $t_2=-0.395$, $t_3=-1.25$, 
$t_4=-0.125$, $t_5=-0.150$~\cite{ref:Comment1}, 
to reproduce the Fermi surface topology. 
The chemical potentials $\mu_{\ell}$'s are determined 
by the condition $n_{\ell}=0.667$, 
where $n_{\ell}$ is the electron number 
of orbital $\ell$ per one spin state 
(The orbitals are evenly filled with electron). 
We introduce the Coulomb interaction part: 
\begin{eqnarray}
H' &=& \frac{U}{2} \sum_{i}\sum_{\ell}
\sum_{\sigma \neq \sigma'} 
c_{i\ell\sigma}^{\dag} c_{i\ell\sigma'}^{\dag} 
c_{i\ell\sigma'} c_{i\ell\sigma} \nonumber\\
&&+ \frac{U'}{2} \sum_{i}\sum_{\ell \neq \ell'}
\sum_{\sigma,\sigma'} 
c_{i\ell\sigma}^{\dag} c_{i\ell'\sigma'}^{\dag} 
c_{i\ell'\sigma'} c_{i\ell\sigma} \nonumber\\
&&+ \frac{J}{2} \sum_{i}\sum_{\ell \neq \ell'}
\sum_{\sigma,\sigma'} 
c_{i\ell\sigma}^{\dag} c_{i\ell'\sigma'}^{\dag} 
c_{i\ell\sigma'} c_{i\ell'\sigma} \nonumber\\
&& + \frac{J'}{2} \sum_{i}\sum_{\ell \neq \ell'}
\sum_{\sigma \neq \sigma'} 
c_{i\ell\sigma}^{\dag} c_{i\ell\sigma'}^{\dag} 
c_{i\ell'\sigma'} c_{i\ell'\sigma}, 
\end{eqnarray}
where the operator $c_{i\ell\sigma}$[$c_{i\ell\sigma}^{\dag}$] 
is the electron annihilation[creation] operator at $i$-th Ru site 
($c_{\mib{k}\ell\sigma}$ is the Fourier transform of $c_{i\ell\sigma}$). 
The microscopic origin of $H'$ is the Coulomb interaction 
between the Ru4d electrons. 
The total Hamiltonian is $H = H_0+H'$. 

As we will see in the next section, the Fermi surface consists 
of three branches (which are named $\alpha$, $\beta$ and $\gamma$), 
consistent with the de Haas-van Alphen oscillation~\cite{ref:Mackenzie1996} 
and photoemission measurements~\cite{ref:Damascelli2000}. 
The Hamiltonian $H_0$ is diagonalized easily, 
and the obtained dispersions are 
\begin{eqnarray}
\xi_{\alpha}(\mib{k}) &=& \xi_+(\mib{k}) 
- \sqrt{\xi_-^2(\mib{k}) + \lambda^2(\mib{k})},\\
\xi_{\beta}(\mib{k}) &=& \xi_+(\mib{k}) 
+ \sqrt{\xi_-^2(\mib{k}) + \lambda^2(\mib{k})},\\
\xi_{\gamma}(\mib{k}) &=& \xi_{xy}(\mib{k}), 
\end{eqnarray}
where $\xi_{\pm}(\mib{k}) 
= (\xi_{yz}(\mib{k}) \pm \xi_{xz}(\mib{k}))/2$. 
The elements of diagonalization matrix are 
\begin{equation}
U(\mib{k})=
\begin{array}{r@{}l}
& \begin{array}{ccc}
\makebox[0.6em]{}\gamma &\makebox[1.1em]{}\alpha &\makebox[1.9em]{}\beta
\end{array} \\
\begin{array}{l}
{\it xy}\\ {\it yz}\\ {\it xz}\\
\end{array} & \left[
\begin{array}{ccc}
1 & 0 & 0 \\
0 & K(\mib{k}) & L(\mib{k})\\
0 & -L(\mib{k}) & K(\mib{k})
\end{array} \right],
\end{array}
\label{eq:MatrixU}
\end{equation}
where 
$K(\mib{k}) = \sqrt{(1-M(\mib{k}))/2}$, \\
$L(\mib{k}) = {\rm sgn}(\lambda(\mib{k}))
\sqrt{(1+M(\mib{k}))/2}$, and 
$M(\mib{k}) = \xi_-(\mib{k})/\sqrt{\xi_-^2(\mib{k})+\lambda^2(\mib{k})}$. 
We define the bare Green's function: 
\begin{equation} 
G_a^{(0)}(k) = ({\rm i} \omega_n - \xi_a(\mib{k}))^{-1}, 
\label{eq:BareGreensFunction}
\end{equation}
the band index $a$ takes $\alpha$, $\beta$ and $\gamma$. 

The anomalous self-energy (i.e., superconducting order parameter) 
is obtained by solving the linearized Eliashberg equation: 
\begin{flushleft}
\begin{eqnarray}
\Sigma_{a,\sigma_1\sigma_2}^{\rm A}(k) = 
- \frac{T}{N} \sum_{a',k',\sigma_3\sigma_4} 
V_{a\sigma_1\sigma_2,a'\sigma_3\sigma_4}(k,k') \nonumber \\
\times |G_{a'}^{(0)}(k')|^2 
\Sigma_{a',\sigma_4\sigma_3}^{\rm A}(k'), 
\label{eq:Eliashberg}
\end{eqnarray}
\end{flushleft}
where $k^({}'{}^)=(\mib{k}{}^({}'{}^),{\rm i}\omega_n{}^({}'{}^))$ 
[$\omega_n = (2n+1) \pi T$], 
and band index $a^({}'{}^)$ takes 
$\alpha$, $\beta$ and $\gamma$. 
The effective interaction 
$V_{a\sigma_1\sigma_2,a'\sigma_3\sigma_4}(k,k')$ 
is evaluated by the third order perturbation expansion in $H'$. 
In the present calculation, 
we take $U=4.0$, $U'=0.33U$, $J=0.33U$, $J'=0.33U$. 
The lengthy procedure of expansions 
and the numerical solutions 
of the Eliashberg equation~(\ref{eq:Eliashberg}) 
are given in the previous work~\cite{ref:Nomura2002b}. 
For the spin-triplet states, 
the order parameter is expressed using the vectorial notation: 
$\Sigma_{a,\sigma_1\sigma_2}^{\rm A}(k) = [{\rm i} (\mib{D}_a(k) 
\cdot \mib{\sigma})\sigma_y]_{\sigma_1\sigma_2}$. 

We assume that the momentum and temperature dependence 
of the superconducting gap function 
$\Delta_a(\mib{k})$ are given 
by the following equation (as in ref.~\ref{ref:Nomura2002a}), 
\begin{equation}
\Delta_a(\mib{k}) = f_a(\mib{k})\Delta(T), 
\end{equation}
where $f_a(\mib{k}) = {\rm Const.} \times D_a(k)$, and 
$D_a(k)$ is the component of the vector $\mib{D}_a(k)$ 
(we assume $\mib{D}_a(k) = D_a(k) \hat{z}$.), 
and obtained by solving the above 
Eliashberg equation~(\ref{eq:Eliashberg}). 
The temperature dependence of the gap magnitude 
$\Delta(T)$ is determined by the standard BCS gap equation: 
\begin{equation}
\Delta_a(\mib{k}) = - \frac{1}{N} \sum_{a',\mib{k}'} 
V_{\mib{k}a,\mib{k}'a'} \frac{\tanh [E_{a'}(\mib{k}')/(2T)]}{E_{a'}(\mib{k}')} 
\Delta_{a'}(\mib{k}'), 
\end{equation}
with $V_{\mib{k}a,\mib{k}'a'}=-f_a(\mib{k})f_{a'}^*(\mib{k}')$ 
and $E_a(\mib{k})=\sqrt{\xi_a^2(\mib{k})+|\Delta_a(\mib{k})|^2}$. 
$f_a(\mib{k})$ is an odd-parity function, satisfying 
the relation $f_a(\mib{k})=-f_a(-\mib{k})$. 
We assume the chiral state is realized: 
\begin{equation}
f_a(\mib{k}) = f_a^x(\mib{k}) \pm {\rm i} f_a^y(\mib{k}), 
\end{equation}
where $f_a^x(\mib{k})$[$f_a^y(\mib{k})$] is 
a real function possessing 
the $k_x$-like[$k_y$-like] symmetry. 
The superconducting gap structure on band $a$ is obtained by 
the absolute magnitude of $\Delta_a(\mib{k})$, 
\begin{equation}
|\Delta_a(\mib{k})| = |f_a(\mib{k})|\Delta(T) 
= \sqrt{{f_a^x(\mib{k})}^2 + {f_a^y(\mib{k})}^2} \Delta(T). 
\end{equation}

For the details of the discussions in this section, 
one could refer to refs.~\ref{ref:Nomura2002b} 
and \ref{ref:Nomura2002a}. 

\subsection{Self-consistent $T$-matrix approximation 
for non-magnetic impurity scattering}
\label{Sc:SCTMatrixAppr}

Here we consider non-magnetic impurity scattering, 
as preliminaries for the following discussions 
on transport properties: 
\begin{equation}
H_{\rm imp} = \sum_{j \in {\rm imp.}} 
\sum_{\ell\ell'\sigma} u_{\ell\ell'} 
c_{j\ell\sigma}^{\dag} c_{j\ell'\sigma}. 
\end{equation}
``$j \in {\rm imp.}$'' means that the summation is 
performed over the impurity positions. 
The impurities are randomly positioned at Ru sites. 
We assume simply $u_{\ell\ell'}=u \delta_{\ell\ell'}$, 
because of the symmetry properties 
of the Ru4d$\varepsilon$ orbital wave functions. 
In addition, we assume that $u$ is much larger 
than the band width $W$ ($u > 100W$ for 
the numerical calculations), 
i.e., the scattering is almost in the unitarity limit. 

In order to discuss the effect of impurities, 
we take the standard $T$-matrix approximation~\cite{ref:Rammer}. 
The self-energy for the impurity scattering is given by 
\begin{equation}
\Sigma_{\ell\ell'}({\rm i}\omega_n) = c T_{\ell\ell'}({\rm i}\omega_n). 
\label{eq:Sigma}
\end{equation}
$c$ is the impurity concentration, and 
we take $c=10^{-6}$ for the present work. 
$T_{\ell\ell'}({\rm i}\omega_n)$ is the $T$-matrix 
obtained by solving the equation, 
\begin{equation}
T_{\ell\ell'}({\rm i}\omega_n)=u_{\ell\ell'}
+ \frac{1}{N} \sum_{\mib{k}_1,\ell_1\ell_2} u_{\ell\ell_1} 
G_{\ell_1\ell_2}(\mib{k}_1,{\rm i}\omega_n) 
T_{\ell_2\ell'}({\rm i}\omega_n), 
\label{eq:TMtrx}
\end{equation}
where $G_{\ell\ell'}(k)$ is the Green's function. 
We have already omitted the terms containing 
$\sum_{\mib{k}} F_{\ell\ell'}(k)$ 
[$F_{\ell\ell'}(k)$ 
is the anomalous Green's function. See below.], 
which vanish in non-$s$-wave superconducting states~\cite{ref:Mineev}, 
in contrast to the case of $s$-wave superconducting state. 
The impurity self-energy and Green's functions 
are related by the Gorkov equation: 
\begin{eqnarray}
G_a(k) &=& G_a^{(0)}(k) + G_a^{(0)}(k) \Sigma_a(k) G_a(k) \nonumber\\
&& - G_a^{(0)}(k) \Delta_a(\mib{k}) F_a^{\dag}(k), \\
F_a(k) &=& G_a^{(0)}(k) \Sigma_a(k) F_a(k) \nonumber\\
&& + G_a^{(0)}(k) \Delta_a(\mib{k}) G_a(-k), \\
F_a^{\dag}(k) &=& G_a^{(0)}(-k) \Sigma_a(-k) F_a^{\dag}(k) \nonumber\\
&& + G_a^{(0)}(-k) \Delta_a^*(\mib{k}) G_a(k). 
\end{eqnarray}
Using the matrix $U(\mib{k})$ in eq.~(\ref{eq:MatrixU}), 
the orbital indices $\ell{}^({}'{}^)$ 
in $\Sigma_{\ell\ell'}({\rm i}\omega_n)$ 
are converted to band index $a$ by 
\begin{eqnarray} 
\Sigma_a(k) &=& \sum_{\ell\ell'} 
U_{a\ell}^{\dag}(\mib{k})U_{\ell'a}(\mib{k}) 
\Sigma_{\ell\ell'}({\rm i}\omega_n), 
\label{eq:SigmaUU}
\end{eqnarray} 
and the band index $a$ in the Green's functions 
are converted to the orbital indices $\ell{}^({}'{}^)$ by 
\begin{eqnarray}
G_{\ell\ell'}(k) &=& \sum_a
U_{\ell a}(\mib{k})U_{a\ell'}^{\dag}(\mib{k}) G_a(k), 
\label{eq:GUU} \\
F_{\ell\ell'}(k) &=& \sum_a 
U_{\ell a}(\mib{k})U_{\ell' a}(\mib{k}) F_a(k), 
\label{eq:FUU}\\
F_{\ell\ell'}(k) &=& \sum_a 
U_{a\ell}^{\dag}(\mib{k})U_{a\ell'}^{\dag}(\mib{k}) 
F_a^{\dag}(k). 
\label{eq:FDUU}
\end{eqnarray}
Defining the renormalized Green's function by 
\begin{eqnarray}
\bar{G}_a(k) &=& ( {G_a^{(0)}}^{-1}(k) - \Sigma_a(k))^{-1} \nonumber\\
&=& ({\rm i}\omega_n - \xi_a (\mib{k}) - \Sigma_a(k))^{-1}, 
\end{eqnarray}
the above Gorkov's equation is simplified to 
\begin{eqnarray}
G_a(k) &=& \bar{G}_a(k) - \bar{G}_a(k) \Delta_a(\mib{k}) F_a^{\dag}(k), 
\label{eq:GorkovEq1}\\
F_a(k) &=& \bar{G}_a(k) \Delta_a(\mib{k}) G_a(-k), \\
F_a^{\dag}(k) &=& \bar{G}_a(-k) \Delta_a^*(\mib{k}) G_a(k).
\label{eq:GorkovEq2}
\end{eqnarray}
By solving eqs.~(\ref{eq:GorkovEq1})-(\ref{eq:GorkovEq2}), 
the Green's functions are obtained: 
\begin{eqnarray}
G_a(k) &=& \frac{1}{\bar{G}_a^{-1}(k)+\bar{G}_a^*(k)|\Delta_a(\mib{k})|^2},\\
F_a(k) &=& \frac{\Delta_a(\mib{k})}
{|\bar{G}_a(k)|^{-2}+|\Delta_a(\mib{k})|^2},\\
F_a^{\dag}(k) &=& \frac{\Delta_a^*(\mib{k})}
{|\bar{G}_a(k)|^{-2}+|\Delta_a(\mib{k})|^2}. 
\end{eqnarray}
We continue this expression analytically 
to the real frequency axis 
(${\rm i} \omega_n \rightarrow z + {\rm i}0$): 
\begin{align}
G_a^{\rm R}(k) &= \nonumber\\
&\frac{\tilde{z}_a^{{\rm R} *}(-k)-\xi_a(\mib{k})}
{(\tilde{z}_a^{\rm R}(k)-\xi_a(\mib{k}))
(\tilde{z}_a^{{\rm R} *}(-k)-\xi_a(\mib{k}))
+|\Delta_a(\mib{k})|^2},\nonumber \\ \\
F_a^{\rm R}(k) &= \nonumber\\
&\frac{\Delta_a(\mib{k})}
{(\tilde{z}_a^{\rm R}(k)-\xi_a(\mib{k}))
(\tilde{z}_a^{{\rm R} *}(-k)-\xi_a(\mib{k}))
+|\Delta_a(\mib{k})|^2},\nonumber \\ \\
F_a^{\dag {\rm R}}(k) &= \nonumber\\
&\frac{ \Delta_a^*(\mib{k})}
{(\tilde{z}_a^{\rm R}(k)-\xi_a(\mib{k}))
(\tilde{z}_a^{{\rm R} *}(-k)-\xi_a(\mib{k}))
+|\Delta_a(\mib{k})|^2}, \nonumber \\
\end{align}
with 
\begin{equation}
\tilde{z}_a^{\rm R}(k) = z - \Sigma_a^{\rm R}(k), 
\label{eq:ztilde}
\end{equation}
and $k = (\mib{k}, z)$. 
Here we may assume the electron-hole symmetry in order 
to simplify the discussion, 
although the actual electronic structure does not 
possess the symmetry. 
This simplification does not affect the results, 
since the temperatures we consider here 
are enough low, 
compared with the characteristic energy scale of 
the asymmetric structure of the density of states. 
Under the electron-hole symmetry, we have the relation 
$\tilde{z}_a^{{\rm R}*}(-k) = -\tilde{z}_a^{\rm R}(k)$, 
and the expressions are simplified to 
\begin{eqnarray}
G_a^{\rm R}(k) &= & \frac{\tilde{z}_a^{\rm R}(k)+\xi_a(\mib{k})}
{{\tilde{z}_a^{\rm R}(k)}^2-{\xi_a(\mib{k})}^2
-|\Delta_a(\mib{k})|^2}, 
\label{eq:GFunction}\\
F_a^{\rm R}(k) &= & -\frac{\Delta_a(\mib{k})}
{{\tilde{z}_a^{\rm R}(k)}^2-{\xi_a(\mib{k})}^2
-|\Delta_a(\mib{k})|^2}, 
\label{eq:FFunction}\\
F_a^{\dag {\rm R}}(k) &= &-\frac{ \Delta_a^*(\mib{k})}
{{\tilde{z}_a^{\rm R}(k)}^2-{\xi_a(\mib{k})}^2
-|\Delta_a(\mib{k})|^2}. 
\end{eqnarray} 
We determine $\tilde{z}_a^{\rm R}(k)$, 
by solving simultaneously 
the analytically continued forms 
of eqs.~(\ref{eq:Sigma}), (\ref{eq:TMtrx}), 
(\ref{eq:SigmaUU}) and (\ref{eq:GUU}), 
and eqs.~(\ref{eq:ztilde}), (\ref{eq:GFunction}). 
There we use $\Delta_a(\mib{k})$ derived 
in the absence of the impurities 
(in \S~\ref{Sc:ModelEliashbergEq}). 
This means that the changes of the gap function 
and its temperature dependence due to the impurities 
are neglected in the present study. 
We consider that the impurity concentration $c$($=10^{-6}$)
is enough small not to change significantly 
the gap structure and its temperature dependence. 

In the present work, we assume that the damping 
of thermally excited quasi-particles is dominantly 
caused by the impurity scattering, and ignore 
the damping effect due to electron-electron scattering. 
We could justify the assumption as follows. 
Within a simple discussion, we expect the normal-state 
electronic thermal conductivity $\kappa_{\rm n}$ 
behaves at low temperatures as 
\begin{equation} 
\kappa^{-1}_{\rm n} \sim d T^{-1} + f T + g T^2, 
\label{eq:NormalThermalCond}
\end{equation}
where $T$ is temperature, and $d$, $f$ and $g$ 
are constants~\cite{ref:Abrikosov}. 
The first, second and third terms of the right hand side 
in eq.~(\ref{eq:NormalThermalCond}) 
originate from the impurity scattering, 
the electron-electron scattering 
and the electron-phonon scattering, respectively. 
According to the experimental data of 
the normal-state thermal conductivity 
in ref.~\ref{ref:Tanatar2001b}, 
we find that the first term is sufficiently 
larger than the other two terms 
within the low-temperature region. 

\subsection{Ultrasound attenuation rate}
\label{Sc:SoundAttenuationRate}

Scattering of phonons by the electron system 
causes the ultrasound attenuation. 
We consider the electron-phonon interaction: 
\begin{equation}
H_{\rm ep} = N^{-\frac{1}{2}} \sum_{\mib{k}\mib{q},\ell\ell',\sigma} 
\Lambda_{\mib{k},\mib{q},\ell\ell'} \phi_{\mib{q}}
c_{\mib{k}+\mib{q}\ell\sigma}^{\dag} c_{\mib{k} \ell'\sigma}, 
\label{eq:ElPhInt}
\end{equation}
where $\phi_{\mib{q}}=b_{\mib{q}} + b_{\mib{-q}}^{\dag}$, 
and $b_{\mib{q}}$($b_{\mib{q}}^{\dag}$) is the 
phonon annihilation(creation) operator with momentum $\mib{q}$. 
The matrix elements $\Lambda_{\mib{k},\mib{q},\ell\ell'}$ 
for the present electronic structure of Sr$_2$RuO$_4$ are given by 
\begin{eqnarray}
\Lambda_{\mib{k},\mib{q},xy,xy} &= &
{\rm i}\{ \tilde{g}_1(\cos k_x \hat{e}_x \hat{q}_x 
+ \cos k_y \hat{e}_y \hat{q}_y) \nonumber\\
&&+ \tilde{g}_2 \cos k_x \cos k_y 
(\hat{e}_x \hat{q}_x + \hat{e}_y \hat{q}_y) \nonumber\\
&&-\tilde{g}_2 \sin k_x \sin k_y (\hat{e}_x \hat{q}_y 
+ \hat{e}_y \hat{q}_x) \}, 
\label{eq:ElPhMtrx1}\\
\Lambda_{\mib{k},\mib{q},yz,yz} &= &
{\rm i}( \tilde{g}_4 \cos k_x \hat{e}_x \hat{q}_x 
+ \tilde{g}_3 \cos k_y \hat{e}_y \hat{q}_y), \\
\Lambda_{\mib{k},\mib{q},xz,xz} &= &
{\rm i}( \tilde{g}_3 \cos k_x \hat{e}_x \hat{q}_x 
+ \tilde{g}_4 \cos k_y \hat{e}_y \hat{q}_y), \\
\Lambda_{\mib{k},\mib{q},yz,xz} &= & \Lambda_{\mib{k},\mib{q},xz,yz} \nonumber\\
&= & 
{\rm i} \tilde{g}_5 \{ - \sin k_x \sin k_y (\hat{e}_x \hat{q}_x 
+ \hat{e}_y \hat{q}_y) \nonumber\\
&&+ \cos k_x \cos k_y (\hat{e}_x \hat{q}_y + \hat{e}_y \hat{q}_x) \},
\label{eq:ElPhMtrx2}
\end{eqnarray}
and the other elements are zero 
(See Appendix~\ref{Ap:ElectronPhononMtrx}). 
$\hat{\mib{e}}=(\hat{e}_x,\hat{e}_y)$ 
and $\hat{\mib{q}}=(\hat{q}_x,\hat{q}_y)$ 
are the unit vectors along the directions 
of phonon polarization and phonon propagation, 
respectively. 

\begin{figure}
\begin{center}
\includegraphics[width=0.7\linewidth]{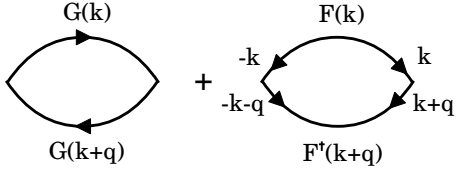}
\end{center}
\caption{The diagrammatic representation 
of the phonon self-energy part 
within the mean field approximation. 
The solid lines with arrowheads denote the Green's functions.} 
\label{Fig:PhononSelfEnergy}
\end{figure}

The energy spectrum of phonon is given 
by the poles of phonon Green's function. 
The phonon Green's function $D(q)$ 
[$q=(\mib{q}, {\rm i}\Omega_m)$, $\Omega_m=2 m \pi T$] 
satisfies 
\begin{equation}
D(q) = D^{(0)}(q) + D^{(0)}(q) \Pi (q) D(q), 
\end{equation}
where $D^{(0)}(q)$ is the bare phonon Green's function 
[$D^{(0)}(q)=-2 \omega_0(\mib{q}) 
/ ({\Omega_m}^2 + {\omega_0}(\mib{q})^2)$], and 
$\Pi (q)$ is the phonon self-energy. 
Using the retarded phonon Green's function $D^R(q)$ 
[$D^R(q)$ is the analytic continuation of $D(q)$ to the real axis 
by ${\rm i} \Omega_m \to \Omega + {\rm i} 0$ $(m>0)$], 
the attenuation rate $\alpha$ of the phonon 
with  momentum $\mib{q}$ is obtained by solving 
${D^R(\mib{q},\Omega)}^{-1} = 0$ for $\Omega$ 
and then taking the imaginary part of the solution: 
$\alpha \equiv {\rm Im} \Omega 
\approx - {\rm Im} \Pi^R(\mib{q},\omega_0(\mib{q}))$. 
$\Pi^R(\mib{q},\Omega)$ is the analytic continuation 
of $\Pi(\mib{q},{\rm i}\Omega_m)$. 
The phonon self-energy $\Pi(q)$ 
within the mean field theory is represented 
by the diagrams in Fig.~\ref{Fig:PhononSelfEnergy}. 
The analytic expression of the phonon self-energy is 
\begin{eqnarray}
\Pi (q) = 2 \frac{T}{N} \sum_{k,\ell_i} 
\Lambda_{\mib{k}, -\mib{q}, \ell_1\ell_2}
\Lambda_{\mib{k}, \mib{q}, \ell_3\ell_4} \nonumber\\
\times
[G_{\ell_2\ell_3}(k+q) G_{\ell_4\ell_1}(k) 
-F_{\ell_1\ell_3}^{\dag}(k+q) F_{\ell_4\ell_2}(k)], 
\end{eqnarray}
where we have used 
the relation $\Lambda_{-\mib{k}, \mib{q}, \ell\ell'}
=\Lambda_{\mib{k}, \mib{q}, \ell\ell'}$, 
and the approximate relation 
$\Lambda_{\mib{k}+\nu \mib{q}, \mib{q}, \ell\ell'} 
\approx \Lambda_{\mib{k}, \mib{q}, \ell\ell'}$ 
(for $\nu \sim 1$) within the precision to the leading order 
of $|\mib{q}|$. The front factor two originates 
from the summation with respect to spin indices. 
We continue $\Pi (q)$ analytically to the real axis, convert 
orbital indices $\ell_i$ to band index $a$, 
and take the imaginary part. 
Since we consider the hydrodynamic limit 
($\ell_{\rm e} \ll 2 \pi / |\mib{q}|$, $\ell_{\rm e}$: mean free path)
in the present study, 
we may take the limit $\mib{q}, \Omega \rightarrow 0$ 
in the argument of the Green's functions. 
Thus we obtain the expression for the attenuation rate $\alpha$: 
\begin{eqnarray}
\alpha = \frac{\omega_0(\mib{q})}{2 \pi T} 
\sum_{a,\mib{k}} |\Lambda_{\mib{k}, \mib{q}, a}|^2 
\int_{- \infty}^{\infty} {\rm d} z \frac{1}{\cosh ^2 (\frac{z}{2T})} \nonumber\\ 
\times \bigl[ ({\rm Im} G_a^{\rm R}(\mib{k}, z))^2 
- |\Delta_a(\mib{k})|^2({\rm Im} \tilde{F}_a^{\rm R}(\mib{k}, z))^2 \bigr], 
\label{eq:AttenuationRate1}
\end{eqnarray}
where $\tilde{F}_a^{\rm R}(\mib{k}, z)$ is defined 
by $F_a^{\rm R}(\mib{k}, z) = \Delta_a(\mib{k}) 
\tilde{F}_a^{\rm R}(\mib{k}, z)$, 
and we have neglected the inter-band crossing terms, 
${\rm Im}G_a(k) {\rm Im} G_{a'}(k)$ and ${\rm Im}F_a(k) 
{\rm Im}F_{a'}(k)$ ($a \neq a'$). 
The contribution from these crossing terms 
is considered to be negligibly small 
in the present case of the long-wavelength limit 
and low temperatures. 
$\Lambda_{\mib{k}, \mib{q}, a}$ is related 
to $\Lambda_{\mib{k}, \mib{q}, \ell\ell'}$ by 
\begin{equation}
\Lambda_{\mib{k}, \mib{q}, a} = \sum_{\ell_1\ell_2} 
U_{a\ell_1}^{\dag}(\mib{k}) U_{\ell_2 a}(\mib{k}) 
\Lambda_{\mib{k}, \mib{q}, \ell_1\ell_2}. 
\end{equation}

Here we perform the momentum integration perpendicular 
to the Fermi surfaces in eq.~(\ref{eq:AttenuationRate1}). 
Using the expressions of the Green's functions 
(\ref{eq:GFunction}) and (\ref{eq:FFunction}), and 
adopting the approximate form of integration measure, 
\begin{equation}
\sum_{a,\mib{k}} \cdots \rightarrow 
\sum_{a,\mib{k}_{\rm F}} 
\biggl|\frac{\partial \xi_a(\mib{k})}
{\partial \mib{k}}\biggr|^{-1}_{{\mib k}={\mib k}_{\rm F}}
\int_{-\infty}^{\infty} {\rm d} \xi_a \cdots, 
\label{eq:ApprIntegrMeasure}
\end{equation}
the formula (\ref{eq:AttenuationRate1}) is reduced to 
\begin{eqnarray}
\alpha = \frac{\omega_0(\mib{q})}{2 \pi T} 
\sum_{a,\mib{k}_{\rm F}} 
|\Lambda_{\mib{k}_{\rm F}, \mib{q}, a}|^2 
\biggl|\frac{\partial \xi_a(\mib{k})}
{\partial \mib{k}}\biggr|^{-1}_{{\mib k}={\mib k}_{\rm F}} 
\int_{- \infty}^{\infty} {\rm d} z 
\int_{-\infty}^{\infty} {\rm d} \xi_a \nonumber\\
\times \frac{1}{\cosh ^2 (\frac{z}{2T})} 
\biggl[ \biggl({\rm Im} 
\frac{\tilde{z}_a^{\rm R}(\mib{k}_{\rm F},z)+\xi_a}
{{\tilde{z}_a^{\rm R}(\mib{k}_{\rm F},z)}^2-{\xi_a}^2
-|\Delta_a(\mib{k}_{\rm F})|^2} \biggr)^2 \nonumber\\
- \biggl({\rm Im} \frac{|\Delta_a(\mib{k}_{\rm F})|}
{{\tilde{z}_a^{\rm R}(\mib{k}_{\rm F},z)}^2-{\xi_a}^2
-|\Delta_a(\mib{k}_{\rm F})|^2}\biggr)^2 \biggr].
\nonumber\\
\label{eq:AttenuationRate2}
\end{eqnarray}
$\sum_{\mib{k}_{\rm F}}$ means the momentum integration 
on the Fermi surface. 
We obtain the final expression of $\alpha$ 
by performing the integration with respect to $\xi_a$: 
\begin{equation}
\alpha = \frac{\omega_0(\mib{q})}{8 T} 
\sum_{a,\mib{k}_{\rm F}} 
|\Lambda_{\mib{k}_{\rm F}, \mib{q}, a}|^2 
\int_{- \infty}^{\infty} {\rm d} z 
\frac{1}{\cosh ^2 (\frac{z}{2T})} I_a(\mib{k}_{\rm F},z), 
\label{eq:AttenuationRate3}
\end{equation}
with 
\begin{eqnarray}
I_a(\mib{k}_{\rm F},z) &=& 
\biggl|\frac{\partial \xi_a(\mib{k})} 
{\partial \mib{k}}\biggr|^{-1}_{{\mib k}={\mib k}_{\rm F}} \nonumber\\
&\times & 
\frac{1}{{\rm Im} 
\sqrt{{\tilde{z}_a^{\rm R}(\mib{k}_{\rm F},z)}^2- 
|\Delta_a(\mib{k}_{\rm F})|^2}}  \nonumber\\
&\times &
\biggl( 1+\frac{|\tilde{z}_a^{\rm R}(\mib{k}_{\rm F},z)|^2 
- |\Delta_a(\mib{k}_{\rm F})|^2}
{|{\tilde{z}_a^{\rm R}(\mib{k}_{\rm F},z)}^2 
- |\Delta_a(\mib{k}_{\rm F})|^2|} \biggr). 
\label{eq:Ia(k)}
\end{eqnarray}

The electron-phonon coupling 
$\Lambda_{\mib{k}, \mib{q}, a}$ must be renormalized 
to satisfy the local charge neutrality condition 
(See Appendix~\ref{Ap:ChargeNeutCondition}, 
and also ref.~\ref{ref:Abrikosov}): 
\begin{equation}
\Lambda_{\mib{k}, \mib{q}, a} \rightarrow 
\Lambda_{\mib{k}, \mib{q}, a}- \bar{\Lambda}_{\mib{q}}, 
\end{equation}
where $\bar{\Lambda}_{\mib{q}}$ is the mean value 
of $\Lambda_{\mib{k}, \mib{q}, a}$ over the Fermi surfaces, 
and given by 
\begin{eqnarray}
\bar{\Lambda}_{\mib{q}} &= & \frac{\sum_{a,\mib{k}} 
\{(\partial_{\xi}G)_a(\mib{k})\} \Lambda_{\mib{k}, \mib{q}, a}}
{\sum_{a,\mib{k}} (\partial_{\xi}G)_a(\mib{k})} \nonumber\\
&= &\frac{\sum_{a,\mib{k},\ell_1\ell_2} 
\{(\partial_{\xi}G)_a(\mib{k})\} U_{a\ell_1}^{\dag}(\mib{k}) 
U_{\ell_2 a}(\mib{k}) \Lambda_{\mib{k}, \mib{q}, \ell_1\ell_2}}
{\sum_{a,\mib{k}} (\partial_{\xi}G)_a(\mib{k})}, \nonumber\\
\label{eq:LambdaMeanValue}
\end{eqnarray}
and 
\begin{eqnarray}
(\partial_{\xi}G)_a(\mib{k}) &= & 
\frac{{\Delta_a(\mib{k})}^2}{2{E_a(\mib{k})}^3} \tanh 
\biggl(\frac{E_a(\mib{k})}{2T} \biggr) \nonumber\\
&&+ \frac{{\xi_a(\mib{k})}^2}{4T{E_a(\mib{k})}^2}
\frac{1}{\cosh^2 (\frac{E_a(\mib{k})}{2T}) }. 
\label{eq:DerivativeG1}
\end{eqnarray}
Throughout the study of sound attenuation, 
we always retain the condition, 
and use the renormalized $\Lambda_{\mib{k}, \mib{q}, a}$. 

\subsection{Thermal conductivity}

The thermal conductivity tensor 
is calculated employing 
Kubo formula~\cite{ref:Mineev,ref:Langer1962,ref:Luttinger1964}: 
\begin{equation}
\kappa_{\mu\nu} = \frac{1}{T} 
\lim_{\Omega \rightarrow 0} \frac{1}{\Omega}
\lim_{\mib{q} \rightarrow 0} 
{\rm Im} P_{\mu\nu}^{\rm R}(\mib{q},\Omega), 
\label{eq:ThermalCond1}
\end{equation}
where $P_{\mu\nu}^{\rm R}(\mib{q},\Omega)$ is 
the retarded thermal-flux correlation function, 
and obtained by continuing analytically 
$P_{\mu\nu}(\mib{q},{\rm i}\Omega_m)$. 
$P_{\mu\nu}(\mib{q},{\rm i}\Omega_m)$ is 
the Fourier transform of 
\begin{equation}
P_{\mu\nu}(\mib{q},\tau) 
= \sum_{a} <T_{\tau} j_{a\mu}^{T}(\mib{q},\tau) 
j_{a\nu}^{T}(-\mib{q},0)>, 
\end{equation}
where $j_{a\mu}^{T}(\mib{q},\tau)$ is the imaginary-time 
thermal-flux operator on band $a$: 
\begin{eqnarray}
j_{a\mu}^{T}(\mib{q},\tau) &=& 
\lim_{\tau' \rightarrow \tau} 
\frac{1}{2} \sum_{\mib{k},\sigma} 
\biggl(\frac{\partial}{\partial \tau} v_{\mib{k+q}a\mu}
-  v_{\mib{k}a\mu} \frac{\partial}{\partial \tau'}\biggr) \nonumber \\
&&\times 
c_{\mib{k}a\sigma}^{\dag}(\tau) c_{\mib{k+q}a\sigma}(\tau'),  
\end{eqnarray}
and $v_{\mib{k}a\mu}$ is the $\mu$-component 
of the velocity on band $a$: 
\begin{equation}
v_{\mib{k}a\mu} = 
\frac{\partial \xi_a(\mib{k})}{\partial k_{\mu}}.
\end{equation} 
Within the mean field theory, the correlation function 
$P(\mib{q},{\rm i}\Omega_m)$ is expanded using 
the Green's functions as 
\begin{eqnarray}
P_{\mu\nu}(\mib{q},{\rm i}\Omega_m) = 
\frac{1}{2} \frac{T}{N} \sum_{a,k} 
\{(\omega_n+\Omega_m) v_{\mib{k}a\mu} 
+ \omega_n v_{\mib{k}+\mib{q}a\mu}\} \nonumber\\
\times 
\{(\omega_n+\Omega_m) v_{\mib{k}a\nu} 
+ \omega_n v_{\mib{k}+\mib{q}a\nu}\} \nonumber\\
\times 
[G_a(k+q) G_a(k) - F_a^{\dag}(k+q) F_a(k)]. 
\label{eq:PFunction}
\end{eqnarray}
We perform the analytic continuation 
${\rm i}\Omega_m \rightarrow \Omega+{\rm i}0$ 
of eq.~(\ref{eq:PFunction}), 
and substitute it into eq.~(\ref{eq:ThermalCond1}), 
then we obtain the expression 
for the thermal conductivity $\kappa_{\mu\nu}$: 
\begin{eqnarray}
\kappa_{\mu\nu} = 
\frac{1}{2 \pi T^2} \sum_{a,\mib{k}} 
v_{\mib{k}a\mu} v_{\mib{k}a\nu} 
\int_{- \infty}^{\infty} {\rm d} z 
\frac{z^2}{\cosh ^2 (\frac{z}{2T})} \nonumber\\ 
\times \bigl[ ({\rm Im} G_a^{\rm R}(\mib{k}, z))^2 
- |\Delta_a(\mib{k})|^2({\rm Im} \tilde{F}_a^{\rm R}(\mib{k}, z))^2 \bigr]. 
\label{eq:ThermalCond2}
\end{eqnarray}
We adopt the approximate integration measure 
in eq.~(\ref{eq:ApprIntegrMeasure}), 
and perform the integration with respect to $\xi_a$, 
as we have done for the sound attenuation rate 
(in \S~\ref{Sc:SoundAttenuationRate}). 
We obtain the final expression for the thermal conductivity: 
\begin{equation}
\kappa_{\mu\nu} =\frac{1}{8 T^2} 
\sum_{a,\mib{k}_{\rm F}} 
v_{\mib{k}_{\rm F}a\mu} v_{\mib{k}_{\rm F}a\nu} 
\int_{- \infty}^{\infty} {\rm d} z 
\frac{z^2}{\cosh ^2 (\frac{z}{2T})} I_a(\mib{k}_{\rm F},z), 
\label{eq:ThermalCond3}
\end{equation}
where $I_a(\mib{k}_{\rm F},z)$ is given by eq.~(\ref{eq:Ia(k)}).

\section{Numerical Analyses of Experimental Data}
\label{Sc:Numerical}

\subsection{Gap structure and density of states}
\label{Sc:GapStructure}

In the present section we provide 
preliminary numerical results: 
gap structure and density of states, 
obtained from the above theoretical formulation. 

\begin{figure}
\begin{center}
\includegraphics[width=0.7\linewidth]{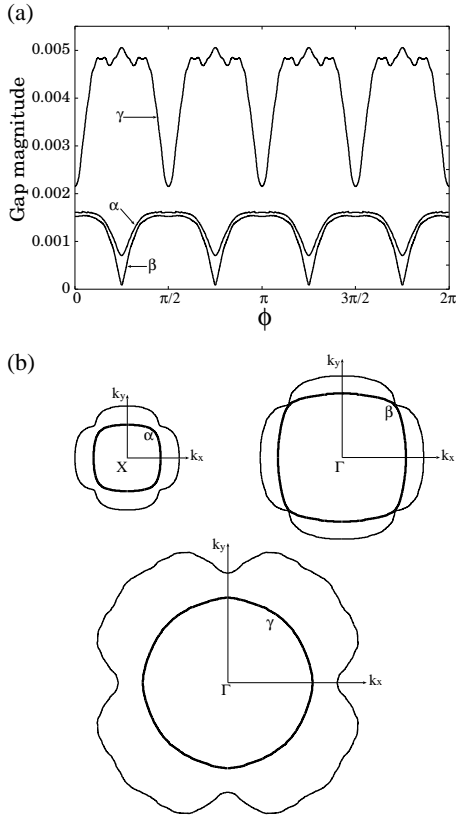}
\end{center}
\caption{
Calculated superconducting gap structure. 
(a) The gap magnitude on each of the three circular(cylindrical)
Fermi surfaces, $\alpha$, $\beta$ and $\gamma$ is depicted 
as a function of the azimuthal angle $\phi$ 
around the $c$-axis (the Ru-O bonding direction 
corresponds to $\phi=0$). 
The unit of energy on the vertical axis 
is about 660(K).
(b) The three Fermi surface sheets and the gap function. 
The Fermi surfaces are depicted by the thick solid lines. 
The dependence of the gap magnitude 
on in-plane direction is expressed 
by the distance from the Fermi circle 
along the direction.} 
\label{Fig:SuperconductingGaps}
\end{figure}

The gap structure calculated by the procedure 
in \S~\ref{Sc:ModelEliashbergEq} is shown in 
Fig.~\ref{Fig:SuperconductingGaps}. 
Since we assume the orbital symmetry $k_x \pm {\rm i} k_y$ 
is realized, the gap function possesses the in-plane fourfold
symmetry (Fig.~\ref{Fig:SuperconductingGaps}). 
We have obtained a gap structure which possesses 
strong in-plane anisotropy and band dependence. 
On the $\gamma$ band, we have a node-like structure 
to the directions $[100]$ and $[010]$. 
This is because of the momentum-space periodicity 
and the odd-parity symmetry of superconducting 
gap function, as pointed out 
by Miyake and Narikiyo~\cite{ref:Miyake1999}. 
On the small-gap bands $\alpha$ and $\beta$, 
nodal structures are obtained 
on the diagonal directions $[110]$ and $[1\bar{1}0]$. 
This is because the $p$-wave attraction 
is weakened around the diagonal points 
by incommensurate antiferromagnetic fluctuation. 
This incommensurate antiferromagnetic fluctuation is 
attributable to the nesting between the Fermi surfaces 
$\alpha$ and $\beta$~\cite{ref:Mazin1999,ref:Nomura2000a,
ref:Eremin2002,ref:Kikugawa2004}. 
Similar nodal structure on the $\alpha$ and $\beta$ bands 
was obtained even by using a two-band model 
for the $\alpha$ and $\beta$ bands~\cite{ref:Kuroki2001}. 

\begin{figure}
\begin{center}
\includegraphics[width=0.7\linewidth]{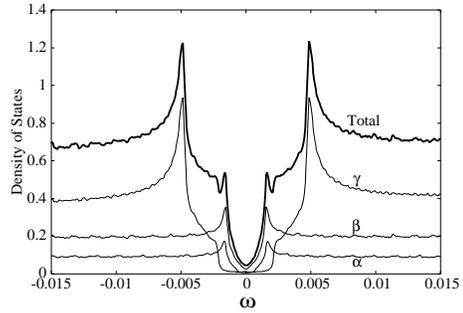}
\end{center}
\caption{
Calculated density of states 
in the superconducting state. 
The vertical axis represents the density of states 
(arb. units) and the horizontal axis represents 
the energy $\omega$ (the Fermi level corresponds 
to $\omega=0$). 
The thick solid line denotes the total density of states, 
and the thin solid lines denote 
the contributions from each band. 
The unit of energy on the horizontal axis 
is about 660(K).} 
\label{Fig:DensityOfStates}
\end{figure}

The density of states (DOS) is calculated by the formula 
\begin{equation}
\rho(\omega) = 
- \frac{1}{\pi} {\rm Im} \frac{1}{N} \sum_{a,\mib{k}} 
G_{a}^{\rm R}(\mib{k}, \omega). 
\end{equation}
The density of states calculated 
for $T=0$ is shown in Fig.~\ref{Fig:DensityOfStates}. 
The main part of the total DOS is taken 
by the main branch $\gamma$. 
The calculated partial DOS is 13.5 \%, 29.1 \% 
and 57.4 \% of the total DOS 
for the $\alpha$, $\beta$ and $\gamma$ bands, respectively. 
This percentage is quantitatively consistent 
with the expectation from the de Haas-van Alphen 
measurement~\cite{ref:Mackenzie1996}. 
The main coherence peaks around $\omega = \pm 0.005$ 
are attributable to the DOS structure on the $\gamma$ band. 
The low energy part of the DOS near the Fermi level 
is dominated by the small-gap bands $\alpha$ and $\beta$. 
Our numerical result predicts that there are 
some fine structures between the main coherence peaks, 
which will originate from the coherence peaks 
on the $\alpha$ and $\beta$ bands. 
Actually, any spectroscopic measurements cannot 
detect such fine peak structures, 
due to their insufficient resolution. 
However, some inflection points might be 
observed in $\rho(\omega)$ as a function 
of energy $\omega$ by spectroscopies. 
Such fine structures, if they are indeed observed 
experimentally, could be considered as the nature 
of the orbital-dependent superconductivity~\cite{ref:Agterberg1997}. 

\subsection{Analysis of ultrasound attenuation rate}
\label{Sc:AnalysisUltrasound}

\begin{figure}
\begin{center}
\includegraphics[width=0.7\linewidth]{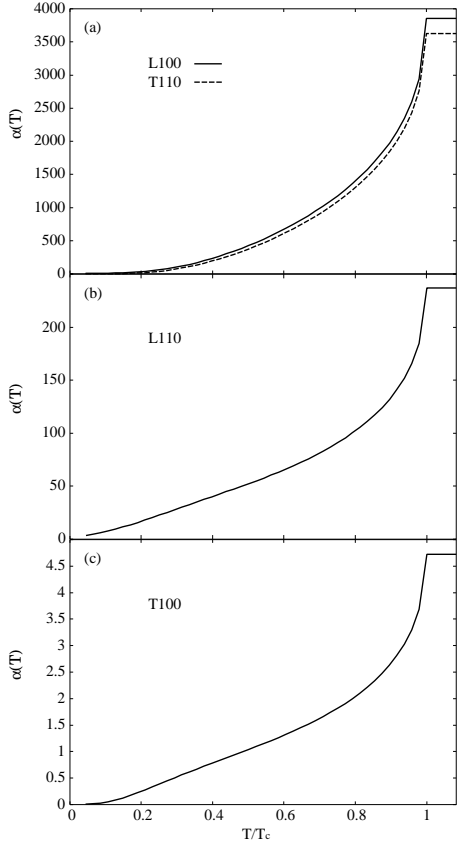}
\end{center}
\caption{
Calculated ultrasound attenuation rate 
for various sound propagation and polarization directions: 
(a) the longitudinal mode propagating along [100] direction 
and the transverse mode along [110] direction 
(i.e., L100: $\hat{\mib{e}} // \hat{\mib{q}}$, $\hat{\mib{q}}//[100]$, 
and  T110: $\hat{\mib{e}} \perp \hat{\mib{q}}$, $\hat{\mib{q}}//[110]$), 
(b) the longitudinal mode along [110] direction 
(i.e., L110: $\hat{\mib{e}} // \hat{\mib{q}}$, $\hat{\mib{q}}//[110]$), 
and (c) the transverse mode along [100] direction 
(i.e., T100: $\hat{\mib{e}} \perp \hat{\mib{q}}$, $\hat{\mib{q}}//[100]$). 
This figure is to be compared with Fig.~2 in ref.~\ref{ref:Lupien2001}. 
}
\label{Fig:AttenRateAll}
\end{figure}

Ultrasound attenuation rate is calculated by using 
the formula (\ref{eq:AttenuationRate3}). 
The electron-phonon coupling matrix 
$\Lambda_{\mib{k}, \mib{q}, \ell \ell'}$ 
is calculated by using 
eqs.~(\ref{eq:ElPhMtrx1})-(\ref{eq:ElPhMtrx2}). 
The constants $\tilde{g}_i$'s are determined 
by fitting to the experimental results 
of ref.~\ref{ref:Lupien2001}. 
In the present study we take 
$\tilde{g}_1= 0.192$, $\tilde{g}_2= 0.0096$, 
$\tilde{g}_3= 0.0672$, $\tilde{g}_4= 0.048$ and 
$\tilde{g}_5= 0.03072$ for $\omega_0(\mib{q})=1.0$. 
The numerical results of ultrasound attenuation rate 
for various propagation and polarization directions 
are shown in Fig.~\ref{Fig:AttenRateAll}. 
The remarkably strong anisotropy is in semiquantitative 
agreement with the experimental results (to be compared 
with Fig.~2 of ref.~\ref{ref:Lupien2001}). 
The attenuation rate of the sound mode L100 
is about one thousand times larger than that of the mode T100. 
As pointed out by Walker and collaborators~\cite{ref:Walker2001}, 
this strong anisotropy is attributable 
to the anisotropy of electron-phonon interaction. 

\begin{figure}
\begin{center}
\includegraphics[width=0.7\linewidth]{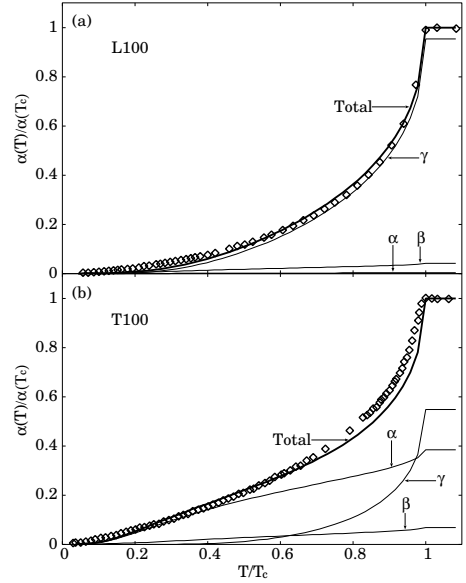}
\end{center}
\caption{
Calculated ultrasound attenuation rates 
(thick solid lines) are compared with the experimental 
data ($\diamond$) read from ref.~\ref{ref:Lupien2001}. 
(a) For the longitudinal wave propagating 
along [100] direction. 
(b) For the transverse wave propagating 
along [100] direction. 
The vertical and horizontal axes represent 
$\alpha(T)$ and temperature $T$ 
normalized by the value at $T=T_{\rm c}$, respectively. 
The contributions from each band are separately shown 
by the thin solid lines.}
\label{Fig:AttenRateWithExp}
\end{figure}

In Fig.~\ref{Fig:AttenRateWithExp}, 
the numerical data are compared 
with the experimental data. 
In addition, the contributions 
from each band are separately presented there 
(note that the quantity is separable 
into the contributions from each band, 
since the formula~(\ref{eq:AttenuationRate3}) 
contains the summation with respect to band index $a$). 

The numerical results are well fitted 
to the experimental results 
in the overall temperature region. 
For L100 mode we can see that the ultrasound 
attenuation is dominated by that on the $\gamma$ band. 
In this case the contributions 
from the $\alpha$ and $\beta$ bands 
are almost negligible. 
On the other hand, for T100 mode, 
the sound attenuation on the $\gamma$ band is ineffective, 
and the contribution from the $\gamma$ band 
is comparable to those from the $\alpha$ and $\beta$ bands. 
The attenuation at low temperatures 
is dominated by the passive bands, 
particularly the $\alpha$ band. 

\subsection{Analysis of thermal conductivity}
\label{Sc:AnalysisThermalCond}

\begin{figure}
\begin{center}
\includegraphics[width=0.7\linewidth]{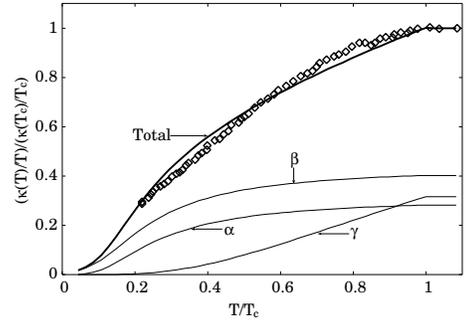}
\end{center}
\caption{
Calculated thermal conductivity $\kappa_{[100]}(T)$ 
(thick solid line) is compared with the experimental 
data ($\diamond$) read from ref.~\ref{ref:Tanatar2001b}. 
The vertical and horizontal axes represent 
$\kappa_{[100]}(T)/T$ and temperature $T$ 
normalized by the value at $T=T_{\rm c}$, respectively. 
The contributions from each band are separately shown 
by the thin solid lines.} 
\label{Fig:ThermalConductivity}
\end{figure}

The thermal conductivity along [100] direction 
(the directions of the thermal current 
and the temperature gradient are both along the $a$-axis)
is calculated using the formula~(\ref{eq:ThermalCond3}). 
After the fitting procedure for specific heat~\cite{ref:Nomura2002a}, 
there remains no fitting parameter, 
in contrast to the case of sound attenuation rate. 
The numerical results are shown 
in Fig.~\ref{Fig:ThermalConductivity}. 
There we can show also the contributions from each band
separately (note that the quantity is separable 
into contributions from each band, 
since the formula~(\ref{eq:ThermalCond3}) 
contains the summation with respect to band index $a$). 

It is very interesting that actually 
the passive bands $\alpha$ and $\beta$ 
contribute significantly to the thermal transport. 
Their contributions are comparable to that 
from the main branch $\gamma$. 
This situation is in contrast 
to other physical quantities in Sr$_2$RuO$_4$. 
For example, the main contribution to the specific heat 
is from the main branch $\gamma$~\cite{ref:Nomura2002a}, 
since the $\gamma$ branch takes the main part 
of the total density of states. 
The reason why the main branch $\gamma$
is not dominant in the thermal transport is as follows. 
The superconducting gap on 
the main branch $\gamma$ has a node-like structure 
near the zone boundaries $(\pm \pi,0)$ and $(0,\pm \pi)$. 
However, the Fermi velocity $\mib{v}_{\mib{k}_{\rm F}}$ 
around these points is quite small, and therefore 
the thermally excited quasi-particles around 
there can not play an essential role for the transport 
(note that the formula~(\ref{eq:ThermalCond3}) 
contains multiplication of the Fermi velocity). 
This situation demonstrates that the thermal transport 
property is inappropriate for detecting 
the gap anisotropy on the $\gamma$ branch. 

\section{Discussion}
\label{Sc:Discussion}

In this section, we suggest some remarks 
from the present calculation. 

As we have seen in \S~\ref{Sc:GapStructure}, 
the gap structure is highly anisotropic. 
Although we have assumed naturally that the chiral state 
$\Delta(\mib{k}) \sim k_x \pm {\rm i} k_y$ 
is realized, the gap function possesses 
sharp depressions on the $\beta$ Fermi surface 
(Fig.~\ref{Fig:SuperconductingGaps}). 
This anisotropy is sufficiently large 
for explaining the power-law temperature dependence 
of various physical quantities. 
We obtain a good fitting for specific heat, 
using the present gap structure, 
as we have shown in the previous work~\cite{ref:Nomura2002a}. 
We should note that this is a new mechanism 
of creating nodal gap structure. 
The incommensurate antiferromagnetic fluctuation 
actually causes the strong gap anisotropy 
on the $\alpha$ and $\beta$ bands. 
Our microscopic theory suggests the following rule 
generally: singlet[triplet] superconducting gap 
function should change[not change] 
its sign between the Fermi surface portions 
which are bridged by antiferromagnetic nesting vector, 
otherwise should take small values (i.e., nodal structure) 
on these Fermi surface portions. 
Recently, Kontani has applied the same scenario 
to nickel-borocarbide superconductors~\cite{ref:Kontani2004}, 
which are considered to be a $s$-wave superconductor 
having a highly anisotropic gap. 
He considered the effect of antiferromagnetic spin fluctuation 
on $s$-wave superconducting gap structure, 
and showed that the gap magnitude is suppressed 
in the Fermi surface regions 
which the antiferromagnetic nesting vector bridges. 
These results suggest that nodal gap structure 
does not always originate from the symmetry properties 
of superconducting order parameter generally. 
So far, there have been a lot of phenomenological discussions. 
They have assumed that superconducting order 
parameter changes its sign on the nodes, 
and have discussed the symmetry properties 
of order parameter on the basis 
of the temperature dependence of physical quantities. 
Such phenomenology has often been applied 
to uranium heavy fermion superconductors~\cite{ref:Machida1999}. 
However, as we have seen above, the validity of 
those phenomenological discussions is not so reliable in general. 

Now we turn our attention to the specific discussions 
on the gap structure of Sr$_2$RuO$_4$. 
Recently Deguchi and Maeno measured specific heat 
under magnetic field~\cite{ref:Deguchi2004a,ref:Deguchi2004b}, 
in order to elucidate the detailed superconducting gap structure. 
They investigated the field-orientation dependence 
of specific heat. 
The anisotropy which is considered to originate 
from the gap anisotropy on the $\gamma$ 
band is observed to vanish at low temperatures. 
This might indicate the possibility that 
the small-gap bands ($\alpha$ and $\beta$) 
possess nodal structures near the diagonal lines, 
and the in-plane anisotropy from the active $\gamma$ band 
is canceled by that from the passive $\alpha$ and $\beta$ 
bands at low temperatures~\cite{ref:Deguchi2004b}. 
Recently Kusunose has analyzed the field-orientation 
dependence of $H_{c_2}$ and specific heat $C$~\cite{ref:Kusunose2004}. 
He showed that the gap structure with the 
intermediate magnitude of minima in $[100]$ direction 
for the $\gamma$ band, and tiny minima of gaps 
in $[110]$ directions for the $\alpha$ and $\beta$ bands 
give consistent behaviors with experiments. 
Particularly, he succeeded in explaining 
the anomalous temperature dependence of the in-plane anisotropy 
of $H_{c_2}$~\cite{ref:Kusunose2004}, i.e., 
the sign change of $\delta H_{c_2}$ 
near $T_{\rm c}$~\cite{ref:Mao2000}. 
These experimental and theoretical results 
are consistent with our gap structure. 
Further investigations should be performed 
toward the thorough clarification 
of the gap structure of Sr$_2$RuO$_4$. 

We have used the electron-phonon coupling 
constants $\tilde{g}_i$'s as fitting parameters. 
It is very difficult to determine 
these parameters microscopically without the fitting. 
In order to determine the parameters $\tilde{g}_i$'s 
without the fitting, it will be indispensable 
to know microscopically how the atoms are rearranged 
in order to relax the local stress (or strain), 
because $\tilde{g}_i$'s will essentially depend 
on the way of the rearrangement. 
For example, the stress along $[100]$ direction 
can be relaxed by reducing the Ru-O-Ru angle 
from 180 degrees without shrinking the Ru-O bonds 
along [100] direction, as well as by shrinking Ru-O bonds 
without changing the Ru-O-Ru angle. 
$\tilde{g}_i$'s will be different 
between these two ways of relaxation. 

We have found actually that the overall temperature dependence 
and anisotropy of the ultrasound attenuation rate 
depend significantly on the parameters $\tilde{g}_i$'s. 
For example, we have shown the results in which 
$\alpha(T)/\alpha(T_{\rm c})$ for L100 is {\it smaller} 
than that for T100 
(Compare Fig.~\ref{Fig:AttenRateWithExp}(a) 
with Fig.~\ref{Fig:AttenRateWithExp}(b)). 
However, we can actually show by using other parameter sets 
of $\tilde{g}_i$'s that $\alpha(T)/\alpha(T_{\rm c})$ for L100 
could be {\it larger} than that for T100. 
Therefore it is actually difficult to determine 
the superconducting gap anisotropy 
by the experimental data of ultrasound attenuation rate 
without any detailed information 
on the electron-phonon coupling matrix elements. 

We find still slight deviation 
between the theoretical and experimental results 
of ultrasound attenuation rate in Fig.~\ref{Fig:AttenRateWithExp}, 
particularly at low temperatures in Fig.~\ref{Fig:AttenRateWithExp}(a). 
We could consider various reasons for the deviation. 
One possible reason is that we could not reproduce 
the Fermi surface deformation precisely 
by using only five components of the electron-phonon 
coupling matrix elements. 
To obtain better agreement, we might require 
higher order harmonics of the matrix elements. 

Recently Contreras {\it et al.} have obtained a good 
fitting of the ultrasound attenuation rate 
in Sr$_2$RuO$_4$~\cite{ref:Contreras2004}. 
Note that their gap structure is quite different from ours. 
They claim that point nodes should be located on the $\gamma$ 
and $\beta$ bands~\cite{ref:Contreras2004}. 
However, it might be difficult to reproduce the 
line-node-like behaviors observed experimentally 
in many quantities. 
In addition, since the jump of specific heat 
at $T_{\rm c}$ will be larger than that in the line-node case, 
it is unclear whether or not to obtain consistency 
with the experiment of specific heat. 
From our microscopic point of view, we expect 
that, since the pairing interaction could 
possess only negligible momentum dependence 
along the $c$-axis due to the strong two-dimensionality, 
point nodes as well as horizontal line nodes 
are almost impossible to realize in Sr$_2$RuO$_4$. 

The field-orientation dependence of thermal conductivity 
has been utilized to investigate the anisotropy 
of superconducting gap~\cite{ref:Izawa2001,
ref:Tanatar2001b,ref:Tanatar2001a}. 
Izawa {\it et al.} observed actually negligible 
in-plane fourfold-symmetry component 
in the field-orientation dependence 
of the thermal conductivity, and concluded 
that line node should run around the cylindrical 
Fermi surfaces (horizontal line node)~\cite{ref:Izawa2001}. 
As we have shown in the present work, 
thermal conductivity is ineffective for 
detecting the gap anisotropy on the $\gamma$ band. 
This will be the reason why the gap anisotropy 
of the $\gamma$ band can be observed 
in field-oriented specific heat~\cite{ref:Deguchi2004a,ref:Deguchi2004b}, 
but not in field-oriented thermal 
conductivity~\cite{ref:Izawa2001,ref:Tanatar2001a}. 
Therefore one might expect that thermal conductivity 
is appropriate for detecting the gap anisotropy 
on the passive bands $\alpha$ and $\beta$. 
However the fourfold-symmetry component 
of the field-orientated thermal conductivity 
may barely be detectable for the $\alpha$ and $\beta$ bands, 
as Tanaka {\it et al.} demonstrated~\cite{ref:Tanaka2003}. 
We consider that the field-oriented thermal conductivity 
does not crucially exclude the possibility of line nodes 
along the $c$-axis. 

In any case, we could conclude that line-node-like structure 
should exist on the $\alpha$ or $\beta$ band, by combining 
the following two points: 
(1) the $T$-linear behavior of $\kappa/T$ 
indicates that line-node-like structure should exist somewhere 
on the Fermi surface and (2) the thermal excitation 
on the $\gamma$ band does not effectively contribute to the thermal transport. 
However, it would still be difficult to obtain some insights 
into the nodal position only from the present analysis 
of thermal conductivity. 

Recently Yanase {\it et al.} proposed another type 
of gap structure~\cite{ref:Yanase}. 
Although the essence of their pairing mechanism 
is the same as ours~\cite{ref:Yanase2003}, 
the gap structure is a little different from ours: 
the nodal position on the $\beta$ band deviates from the diagonals. 
It is still controversial whether 
such a deviation is realized or not. 

For determining the nodal positions thoroughly, 
it would be required to measure the gap magnitude 
by some momentum-resolving experiment 
on the Fermi surface. 
Further experimental probes crucially determining 
the nodal positions are desired to be developed. 

\section{Conclusion}
\label{Sc:Conclusion}

In the present article, we have discussed 
the transport properties in the superconducting state 
of Sr$_2$RuO$_4$, by analyzing the sound attenuation rate 
and thermal conductivity. 
We have shown that the gap structure given 
in Fig.~\ref{Fig:SuperconductingGaps} is consistent 
with the experimental data. 

As we have seen in Fig.~\ref{Fig:SuperconductingGaps}, 
we obtain a nodal gap structure 
with a large in-plane anisotropy, 
particularly on the passive bands 
$\alpha$ and $\beta$, 
even if we assume the chiral state 
$\Delta(\mib{k}) \sim k_x \pm {\rm i} k_y$. 
We should note this is a new mechanism 
of creating nodal gap structure: 
antiferromagnetic fluctuation deforms 
the gap structure, and creates 
the nodal gap structure. 
We do not consider that such a nodal structure 
due to this mechanism occurs exceptionally 
only in Sr$_2$RuO$_4$. 
We consider that a nodal structure 
could occur due to this mechanism 
also in other superconductors, particularly 
likely in strongly correlated electron systems. 
Therefore we should notice generally: 
actually it is not reliable to determine 
pairing symmetry by power-law temperature 
dependence of physical quantities. 

Through the present study, we would like to stress 
that too much simplified models 
(e.g., an isotropic Fermi surface, 
only a single band or a gap function approximated 
by only a few harmonic functions.) 
are insufficient in general to analyze the anisotropy 
and the temperature dependence of physical quantities. 
In addition, we should note in general that some quantities 
can succeed in detecting the anisotropy 
of superconducting gap, but others can not. 
An example is the thermal conductivity in Sr$_2$RuO$_4$, 
which would fail to detect the in-plane gap anisotropy 
on the $\gamma$ band, as we have discussed 
in \S~\ref{Sc:Discussion}. 

In any case, we consider that it is very difficult 
to explain the experimental data without assuming the existence 
of line nodes on the passive small-gap bands $\alpha$ and $\beta$. 

The present work is one of a few examples 
in which physical quantities are analyzed 
with use of realistic many-band electronic 
structure and superconducting gap structure 
obtained by microscopic calculation. 
In conclusion, we hope that such detailed analyses 
are performed also for many other anisotropic 
superconductors.  

\section*{Acknowledgments}
The present work has been developed from a series of works 
performed with Prof. Kosaku Yamada at Kyoto University. 
The author would like to express the deepest gratitude to him. 
It is also a great pleasure for the author 
to thank Prof. Yoshiteru Maeno, Prof. Manfred Sigrist, 
Prof. Kazushige Machida, Dr. Kazuhiko Deguchi, 
and Dr. Hiroaki Kusunose for invaluable communications. 
The numerical calculations in the present work 
were partly performed at the Yukawa Institute Computer Facility 
of Kyoto University. 

\appendix

\section{Electron-Phonon Coupling Matrix}
\label{Ap:ElectronPhononMtrx}

We consider the electron-phonon interaction, 
since the change of electronic structure 
due to lattice distortion is essential 
for the ultrasound attenuation. 
The detailed information about the anisotropy of electron-phonon 
interaction is actually important for the realistic analysis, 
as Walker {\it et al.} pointed out~\cite{ref:Walker2001}. 

The non-interacting part of the Hamiltonian $H_0$ 
is written in the form, 
\begin{equation}
H_0 = \sum_{ii',\ell\ell',\sigma} 
t_{\ell\ell'}[\mib{r}_i-\mib{r}_{i'}] 
c_{i\ell\sigma}^{\dag} c_{i'\ell'\sigma}. 
\end{equation}
$t_{\ell\ell'}[\mib{r}_i-\mib{r}_{i'}]$ is 
the transfer matrix element between the Wannier atomic 
orbitals $\ell$ at $i$-th site and $\ell'$ at $i'$-th site . 
Now the lattice deformation is introduced: 
\begin{equation}
H_0 = \sum_{ii',\ell\ell',\sigma} 
t_{\ell\ell'}[(\mib{r}_i + \mib{u}_i)
-(\mib{r}_{i'} + \mib{u}_{i'})] 
c_{i\ell\sigma}^{\dag} c_{i'\ell'\sigma}. 
\end{equation}
$\mib{u}_i$ is the displacement of the $i$-th Ru atom. 
The hopping integrals are expanded in $\mib{u}_i$: 
\begin{eqnarray}
t_{\ell\ell'}[(\mib{r}_i + \mib{u}_i)-(\mib{r}_{i'} + \mib{u}_{i'})] \nonumber\\
= t_{\ell\ell'}[\mib{r}_i - \mib{r}_{i'}] 
+ \mib{g}_{\ell\ell'}(\mib{r}_i-\mib{r}_{i'}) \cdot (\mib{u}_i- \mib{u}_{i'})
+ ... , 
\end{eqnarray}
where $\mib{g}_{\ell\ell'}(\mib{r}_i-\mib{r}_{i'}) 
= \partial t_{\ell\ell'}[\mib{a}]/ 
\partial \mib{a} |_{\mib{a}=\mib{r}_i-\mib{r}_{i'}}$ 
The Hamiltonian of the electron-phonon interaction is given by 
\begin{equation}
H_{\rm ep} = \sum_{ii',\ell\ell',\sigma} 
\mib{g}_{\ell\ell'}(\mib{r}_i-\mib{r}_{i'}) 
\cdot (\mib{u}_i-\mib{u}_{i'}) 
c_{i\ell\sigma}^{\dag} c_{i'\ell'\sigma}. 
\end{equation}
Then we transform the expression to the momentum representation, 
and apply the second quantization to the lattice vibration, 
\begin{eqnarray}
H_{\rm ep} = \sum_{\mib{R},\ell\ell',\sigma} 
(\mib{g}_{\ell\ell'}(\mib{R}) \cdot \hat{\mib{e}})
\sum_{\mib{q}} (2NM {\omega}_0(\mib{q}))^{-1/2} \nonumber\\
\times
\sum_{\mib{k}} e^{- {\rm i} \mib{k} \cdot \mib{R}}
(1-e^{- {\rm i} \mib{q} \cdot \mib{R}}) (b_{\mib{q}} + b_{-\mib{q}}^{\dag})
c_{\mib{k}+\mib{q}\ell\sigma}^{\dag} c_{\mib{k} \ell'\sigma}.  
\label{eq:Hel-ph}
\end{eqnarray}
Thus the Hamiltonian takes the form 
of eq.~(\ref{eq:ElPhInt}), 
using the electron-phonon coupling matrix 
\begin{eqnarray}
\Lambda_{\mib{k},\mib{q},\ell\ell'} 
&\equiv& (2M {\omega}_0(\mib{q}))^{-1/2} \sum_{\mib{R}} 
e^{- {\rm i} \mib{k} \cdot \mib{R}} \nonumber \\
&&\times(\mib{g}_{\ell\ell'}(\mib{R}) \cdot \hat{\mib{e}})
(1-e^{- {\rm i} \mib{q} \cdot \mib{R}}). 
\end{eqnarray}
$M$ is the ionic mass, $\omega_{\mib{q}}$ is the phonon frequency, 
and $b_{\mib{q}}$($b_{\mib{q}}^{\dag}$) is 
the annihilation(creation) operator of the phonon 
with momentum $\mib{q}$ and polarization $\hat{\mib{e}}$. 
The unit vector $\hat{\mib{e}}$ points to the phonon polarization direction. 
In the actual experiment, the phonon wavenumber is much larger 
than the lattice constant ($\mib{q} \ll 1$), 
while the summation in $\mib{R}$ of eq.~(\ref{eq:Hel-ph})
is performed at most up to the second nearest sites. 
Therefore, $H_{\rm ep}$ is approximated as follows 
by expanding the factor $e^{ - {\rm i} \mib{q} \cdot \mib{R}}$ 
in the power of $\mib{q}$ and neglecting higher orders: 
\begin{equation}
\Lambda_{\mib{k},\mib{q},\ell\ell'} 
= {\rm i} (2M {\omega}_0(\mib{q}))^{-1/2} 
\sum_{\mib{R}} e^{- {\rm i} \mib{k} \cdot \mib{R}} 
(\mib{g}_{\ell\ell'}(\mib{R}) \cdot \hat{\mib{e}})
(\mib{q} \cdot \mib{R}). 
\label{eq:Lambdael-ph}
\end{equation}

Here we consider the specific expression 
of the function $\mib{g}_{\ell\ell'}(\mib{R})$. 
Since we expect naturally that 
the electronic transfers perpendicular 
to the direction of the deformation 
are not significantly affected by the deformation, 
we may assume that $\mib{g}_{\ell\ell'}(\mib{R})$ 
is parallel to the vector $\mib{R}$: 
\begin{equation}
\mib{g}_{\ell\ell'}(\mib{R}) = g_{\ell\ell'}(\mib{R})\hat{\mib{R}}. 
\end{equation}
$\hat{\mib{R}}$ is the unit vector along $\mib{R}$, i.e., 
$\hat{\mib{R}}=\mib{R}/|\mib{R}|$, and 
$g_{\ell\ell'}(\mib{R})$ is an even function of $\mib{R}$ 
[i.e., $g_{\ell\ell'}(\mib{R})=g_{\ell\ell'}(-\mib{R})$]. 
The coefficients $g_{\ell\ell'}(\mib{R})$ are characterized 
by five constants, since we have taken the five hopping 
integrals ($t_1,...,t_5$) for describing 
the electronic structure: 
\begin{eqnarray}
g_{xy,xy}(\hat{\mib{x}}) &=& g_1, 
\label{eq:g1el-ph}\\
g_{xz,xz}(\hat{\mib{x}}) &=& g_3, \\
g_{yz,yz}(\hat{\mib{x}}) &=& g_4, \\
g_{xy,xy}(\hat{\mib{y}}) &=& g_1, \\
g_{xz,xz}(\hat{\mib{y}}) &=& g_4, \\
g_{yz,yz}(\hat{\mib{y}}) &=& g_3, \\
g_{xy,xy}(\hat{\mib{x}}+\hat{\mib{y}}) &=& g_2, \\
g_{yz,xz}(\hat{\mib{x}}+\hat{\mib{y}}), 
&=& g_{xz,yz}(\hat{\mib{x}}+\hat{\mib{y}}) = g_5, \\
g_{xy,xy}(\hat{\mib{x}}-\hat{\mib{y}}) &=& g_2, \\
g_{yz,xz}(\hat{\mib{x}}-\hat{\mib{y}}) 
&=& g_{xz,yz}(\hat{\mib{x}}-\hat{\mib{y}}) = -g_5, 
\label{eq:g2el-ph}
\end{eqnarray}
and the other elements of $g_{\ell\ell'}(\mib{R})$ are zero, 
where $\hat{\mib{x}}$($\hat{\mib{y}}$) is the unit lattice vector 
along the $x$-($y$-)axis. 
We obtain the momentum-dependent electron-phonon coupling matrix elements 
in eqs.~(\ref{eq:ElPhMtrx1})-(\ref{eq:ElPhMtrx2}), 
by substituting eqs.~(\ref{eq:g1el-ph})-(\ref{eq:g2el-ph}) 
into eq.~(\ref{eq:Lambdael-ph}) 
and using the coefficients $\tilde{g}_i$'s 
related to $g_i$'s as 
$\tilde{g}_1 = 2 \gamma g_1$, 
$\tilde{g}_2 = 2^{3/2} \gamma g_2$, 
$\tilde{g}_3 = 2 \gamma g_3$, 
$\tilde{g}_4 = 2 \gamma g_4$, 
$\tilde{g}_5 = 2^{3/2} \gamma g_5$ {}
($ \gamma = \{\omega_0(\mib{q})/(2Mv_{\mib{q}}^2)\}^{1/2}$, 
and $v_{\mib{q}}$ is the sound velocity). 

\section{Charge Neutrality Condition under Lattice Deformation}
\label{Ap:ChargeNeutCondition}

In this Appendix, we consider the local charge neutrality 
condition, which should be retained under the lattice distortion 
due to the ultrasound propagation. 
We introduce a lattice distortion described 
by $\eta \phi_{\mib{q}}$ 
[For the definition of $\phi_{\mib{q}}$, 
see eq.~(\ref{eq:ElPhInt})], 
whose spatial periodicity is characterized 
by the wavenumber $\mib{q}$. 
$\eta$ is related to the magnitude of the displacement 
of the atoms. 
This distortion plays a role of periodic potential  
acting on the electron system. 
The electron self-energy will be modified by 
\begin{equation}
\Sigma_{\ell\ell'}(\mib{k}; q) 
= \eta D(q) \Lambda_{\mib{k},\mib{q},\ell\ell'}. 
\end{equation}
\begin{figure}
\begin{center}
\includegraphics[width=0.7\linewidth]{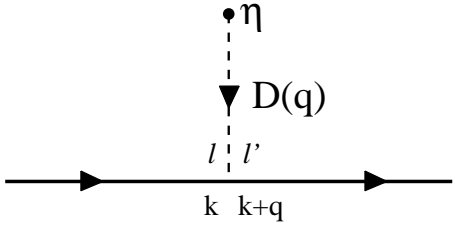}
\end{center}
\caption{The diagrammatic representation 
of the lowest-order electron self-energy correction 
due to the lattice distortion $\eta \phi_{\mib{q}}$.} 
\label{Fig:LatticeDistortionSelfEnergy}
\end{figure}
Now we focus on a local small volume, which is enough smaller 
than the characteristic length of lattice distortion 
($\sim 2 \pi / |\mib{q}|$) but still contains 
macroscopic number of electrons. 
In this volume, the electron system is regarded 
as spatially uniform. 
As far as we discuss the properties of the 
electrons contained in this small volume, 
we can use the Green's function $G_{\ell\ell'}(k)$ corrected 
by the self-energy $\Sigma_{\ell\ell'}(\mib{k}; q)$. 
The local change of electron number is obtained by 
\begin{eqnarray}
\delta n &=& 2 \frac{T}{N}\sum_{k,\ell} 
[ G_{\ell\ell}(k)-G_{\ell\ell}^{(0)}(k) ] \nonumber\\
&\approx& 2 \frac{T}{N}\sum_{a,k} 
\frac{\partial G^{(0)}_a(k)}{\partial \xi_a(\mib{k})}\nonumber\\
&&\times \sum_{\ell_1\ell_2} U_{a\ell_1}^{\dag}(\mib{k}) 
U_{\ell_2 a}(\mib{k}) \Sigma_{\ell_1\ell_2}(\mib{k}; q), 
\end{eqnarray}
within the precision to the lowest order in $\eta$, 
and we have used similar relations 
to eqs.~(\ref{eq:SigmaUU}) and (\ref{eq:GUU}). 
The charge neutrality condition states that the charge inhomogeneity 
caused by the lattice distortion $\eta \phi_{\mib{q}}$ should be canceled 
by introducing a local change of electron chemical 
potential~\cite{ref:Abrikosov}: 
\begin{eqnarray}
\frac{1}{N}\sum_{a,\mib{k}} (\partial_{\xi}G)_a(\mib{k}) 
\biggl[ \sum_{\ell_1\ell_2} 
U_{a\ell_1}^{\dag}(\mib{k}) U_{\ell_2 a}(\mib{k}) 
\Sigma_{\ell_1\ell_2}(\mib{k}; q) \nonumber\\
- \Delta \mu  \biggr] \approx 0, 
\label{eq:ChargeNeutCondition1}
\end{eqnarray}
where $(\partial_{\xi}G)_a(\mib{k})$ is defined as 
\begin{equation}
(\partial_{\xi}G)_a(\mib{k}) 
\equiv T \sum_{\omega_n} 
\frac{\partial G_a^{(0)}(k)}{\partial \xi_a(\mib{k})}. 
\label{eq:DerivativeG2}
\end{equation}
We obtain from eq.~(\ref{eq:ChargeNeutCondition1}) 
\begin{equation}
\Delta \mu = \frac{\sum_{a,\mib{k},\ell_1\ell_2} 
\{(\partial_{\xi}G)_a(\mib{k})\}
U_{a\ell_1}^{\dag}(\mib{k}) 
U_{\ell_2 a}(\mib{k}) \Sigma_{\ell_1\ell_2}(\mib{k}; q)} 
{\sum_{a,\mib{k}} (\partial_{\xi}G)_a(\mib{k}) }. 
\end{equation}

Here let us note that the shift of chemical potential $\Delta \mu$ 
can be regarded as a renormalization of the electron-phonon coupling 
matrix elements $\Lambda_{\mib{k},\mib{q},\ell\ell'}$. 
The chemical potential shift $\Delta \mu$ appears 
in the equations always together with the self-energy 
in the form, $\Sigma_{\ell\ell'}(\mib{k}; q) 
- \Delta \mu \delta_{\ell\ell'}$. 
If we define the mean value of $\Lambda_{\mib{k},\mib{q},\ell\ell'}$ 
by eq.~(\ref{eq:LambdaMeanValue}), 
then 
\begin{equation}
\Sigma_{\ell\ell'}(\mib{k}; q) 
- \Delta \mu \delta_{\ell\ell'} 
= \eta D(q) (\Lambda_{\mib{k},\mib{q},\ell\ell'} 
- \bar{\Lambda}_{\mib{q}} \delta_{\ell\ell'}). 
\end{equation}
Thus the local charge neutrality condition is satisfied 
by using the renormalized value of the electron-phonon coupling, 
$\Lambda_{\mib{k},\mib{q},\ell\ell'} 
- \bar{\Lambda}_{\mib{q}} \delta_{\ell\ell'}$, 
instead of $\Lambda_{\mib{k},\mib{q},\ell\ell'}$. 

In the normal state, using 
eqs.~(\ref{eq:BareGreensFunction}) and (\ref{eq:DerivativeG2}), 
we obtain 
\begin{equation}
(\partial_{\xi}G)_a(\mib{k}) 
= - \frac{\partial n_{\rm F}(\epsilon)}
{\partial \epsilon}\biggr|_{\epsilon=\xi_a(\mib{k})}. 
\end{equation}
$n_{\rm F}(\epsilon)$ is the Fermi function.
In the superconducting state, 
we obtain eq.~(\ref{eq:DerivativeG1}) 
by using eq.~(\ref{eq:DerivativeG2})
and $G_a^{(0)}(k)$ of the form 
\begin{equation}
G_a^{(0)}(k) = - \frac{{\rm i}\omega_n +\xi_a(\mib{k})}
{\omega_n^2  + {\xi_a(\mib{k})}^2 + |\Delta_a(\mib{k})|^2}. 
\end{equation}

\section{
About the Strong Gap Anisotropy on the $\alpha$ and $\beta$ Bands}
\label{Ap:StrongGapAnisotropy}

\begin{figure}
\begin{center}
\includegraphics[width=0.7\linewidth]{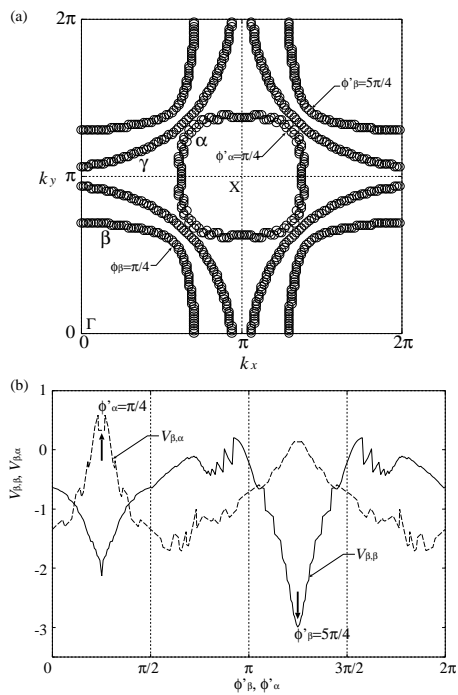}
\end{center}
\caption{(a) The three Fermi surfaces, $\alpha$, $\beta$ and $\gamma$ 
are depicted. Position on the Fermi surfaces $\beta$[$\alpha$] 
is characterized by the in-plane azimuthal angle $\phi$ 
around the points $\Gamma(0,0)$[$X(\pi,\pi)$] 
with respect to the $x$-axis. 
(b) Calculated effective pair-scattering amplitudes 
$V_{\beta,\beta}(\phi_{\beta}=\pi/4;\phi'_{\beta})$ 
and $V_{\beta,\alpha}(\phi_{\beta}=\pi/4;\phi'_{\alpha})$ 
are shown as a function of the azimuthal angles 
$\phi'_{\beta}$ and $\phi'_{\alpha}$, respectively.}
\label{Fig:InteractionAnisotropy}
\end{figure}

In this Appendix, we discuss the origin of the strong gap anisotropy 
on the $\alpha$ and $\beta$ bands by extracting the pairing interaction 
$V_{a\sigma_1\sigma_2,a'\sigma_3\sigma_4}(k,k')$ in eq.~(\ref{eq:Eliashberg}). 
We focus on the values of $V_{\beta\sigma\sigma,\beta\sigma\sigma}(k,k')$ 
and $V_{\beta\sigma\sigma,\alpha\sigma\sigma}(k,k')$ on the Fermi surface. 
We characterize position on the Fermi surface 
by the in-plane azimuthal angle $\phi$ 
with respect to the Ru-O bonding direction. 
Hereafter we refer to the values 
of $V_{\beta\sigma\sigma,\beta\sigma\sigma}(k,k')$ 
and $V_{\beta\sigma\sigma,\alpha\sigma\sigma}(k,k')$ on the Fermi surface 
as $V_{\beta,\beta}(\phi_{\beta};\phi'_{\beta})$ 
and $V_{\beta,\alpha}(\phi_{\beta};\phi'_{\alpha})$, respectively 
(The Matsubara frequencies $\omega_n$ 
and $\omega'_n$ are fixed to $\pi T$). 
Now we fix the initial state of Cooper pair 
on the $\beta$ Fermi surface by $\phi_{\beta} = \frac{\pi}{4}$ 
as pointed in Fig.~\ref{Fig:InteractionAnisotropy}(a). 
$V_{\beta,\beta}(\phi_{\beta}=\pi/4;\phi'_{\beta})$ 
and $V_{\beta,\alpha}(\phi_{\beta}=\pi/4;\phi'_{\alpha})$ 
are depicted as a function of the angles 
$\phi'_{\beta}$ and $\phi'_{\alpha}$, respectively, 
in Fig.~\ref{Fig:InteractionAnisotropy}(b). 
We find that there are characteristic local minimum 
in $V_{\beta,\beta}(\phi_{\beta}=\pi/4;\phi'_{\beta})$ 
around $\phi'_{\beta}=5 \pi/4$ and 
local maximum in $V_{\beta,\alpha}(\phi_{\beta}=\pi/4;\phi'_{\alpha})$ 
around $\phi'_{\alpha}=\pi/4$. 
The local minimum of $V_{\beta,\beta}(\phi_{\beta}=\pi/4;\phi'_{\beta})$ 
around $\phi'_{\beta}=5\pi/4$ favors the same sign of order parameter 
between the points $\phi_{\beta}=\pi/4$ and $\phi'_{\beta}=5\pi/4$. 
The local maximum of $V_{\beta,\alpha}(\phi_{\beta}=\pi/4;\phi'_{\alpha})$ 
around $\phi'_{\alpha}=\pi/4$
favors the different sign of order parameter 
between the points $\phi_{\beta}=\pi/4$ 
and $\phi'_{\alpha}=\pi/4$. 
These momentum dependences deform the gap structure 
on the $\alpha$ and $\beta$ bands significantly, 
and result in the strong anisotropy of gap structure. 
Those characteristic momentum dependences 
of the pairing interaction basically originate 
from those of the fluctuation due to the nesting 
between the $\alpha$ and $\beta$ Fermi surfaces, 
although it is somewhat modulated 
by the diagonalization matrix $U(\mib{k})$ 
in eq.~(\ref{eq:MatrixU}). 


\begin{thebibliography}{99}

\bibitem{ref:Mackenzie2003} 
For a review, A.P. Mackenzie and Y. Maeno: 
Rev. Mod. Phys. {\bf 75} (2003) 657. 

\bibitem{ref:Ishida1998} 
K. Ishida, H. Mukuda, Y. Kitaoka, K. Asayama, 
Z.Q. Mao, Y. Mori and Y. Maeno: 
Nature {\bf 396} (1998) 658. 

\bibitem{ref:Duffy2000} 
J. A. Duffy, S. M. Hayden, Y. Maeno, 
Z. Mao, J. Kulda and G. J. McIntyre: 
Phys. Rev. Lett. {\bf 85} (2000) 5412. 

\bibitem{ref:Rice1995} 
T.M. Rice and M. Sigrist: 
J. Phys.: Condens. Matter {\bf 7} (1995) L643. 

\bibitem{ref:Mazin1997} 
I.I. Mazin and D.J. Singh: 
Phys. Rev. Lett. {\bf 79} (1997) 733. 

\bibitem{ref:Sidis1999} 
Y. Sidis, M. Braden, P. Bourges, B. Hennion, 
S. NishiZaki, Y. Maeno and Y. Mori: 
Phys. Rev. Lett. {\bf 83} (1999) 3320. 

\bibitem{ref:Nomura2000b} 
T. Nomura and K. Yamada: 
J. Phys. Soc. Jpn. {\bf 69} (2000) 3678.
\label{ref:Nomura2000b}

\bibitem{ref:Nomura2002b}
T. Nomura and K. Yamada: 
J. Phys. Soc. Jpn. {\bf 71} (2002) 1993.
\label{ref:Nomura2002b}

\bibitem{ref:YanaseRev2003}
Y. Yanase, T. Jujo, T. Nomura, H. Ikeda, 
T. Hotta and K. Yamada: 
Phys. Rep. {\bf 387} (2003) 1.

\bibitem{ref:Luke1998} 
G.M. Luke, Y. Fudamoto, K.M. Kojima, M.I. Larkin, 
J. Merrin, B. Nachumi, Y.J. Uemura, Y. Maeno, 
Z.Q. Mao, Y. Mori, H. Nakamura and M. Sigrist: 
Nature {\bf 394} (1998) 558. 

\bibitem{ref:NishiZaki2000} 
S. NishiZaki, Y. Maeno and Z.Q. Mao: 
J. Phys. Soc. Jpn. {\bf 69} (2000) 572.

\bibitem{ref:Ishida2000} 
K. Ishida, H. Mukuda, Y. Kitaoka, Z.Q. Mao, 
Y. Mori and Y. Maeno: 
Phys. Rev. Lett. {\bf 84} (2000) 5387.

\bibitem{ref:Bonalde2000} 
I. Bonalde, B.D. Yanoff, M.B. Salamon, 
D.J. Van Harlingen, E.M.E. Chia, Z.Q. Mao 
and Y. Maeno: 
Phys. Rev. Lett. {\bf 85} (2000) 4775.

\bibitem{ref:Lupien2001} 
C. Lupien, W.A. MacFarlane, C. Proust, L. Taillefer, 
Z.Q. Mao and Y. Maeno: 
Phys. Rev. Lett. {\bf 86} (2001) 5986. 
\label{ref:Lupien2001}

\bibitem{ref:Izawa2001} 
K. Izawa, H. Tanaka, H. Yamaguchi, Y. Matsuda, M. Suzuki, 
T. Sasaki, T. Fukase, Y. Yoshoda, R. Settai and Y. Onuki: 
Phys. Rev. Lett. {\bf 86} (2001) 2653. 

\bibitem{ref:Tanatar2001b} 
M.A. Tanatar, S. Nagai, Z.Q. Mao, 
Y. Maeno and T. Ishiguro: 
Phys. Rev. B {\bf 63} (2001) 064505. 
\label{ref:Tanatar2001b}

\bibitem{ref:Agterberg1997} 
D.F. Agterberg, T.M. Rice and M. Sigrist: 
Phys. Rev. Lett. {\bf 78} (1997) 3374. 

\bibitem{ref:Miyake1999} 
K. Miyake and O. Narikiyo: 
Phys. Rev. Lett. {\bf 83} (1999) 1423. 

\bibitem{ref:Hasegawa2000} 
Y. Hasegawa, K. Machida and M. Ozaki: 
J. Phys. Soc. Jpn. {\bf 69} (2000) 336.
\label{ref:Hasegawa2000}

\bibitem{ref:Won2000}
H. Won and K. Maki: 
Europhys. Lett. {\bf 52} (2000) 427. 

\bibitem{ref:Dahm2000} 
T. Dahm, H. Won and K. Maki: 
cond-mat/0006301. 

\bibitem{ref:Zhitomirsky2001} 
M.E. Zhitomirsky and T.M. Rice: 
Phys. Rev. Lett. {\bf 87} (2001) 057001. 

\bibitem{ref:Nomura2002a} 
T. Nomura and K. Yamada: 
J. Phys. Soc. Jpn. {\bf 71} (2002) 404.
\label{ref:Nomura2002a}

\bibitem{ref:Moreno1996} 
J. Moreno and P. Coleman: 
Phys. Rev. B {\bf 53} (1996) R2995. 
\label{ref:Moreno1996}

\bibitem{ref:Graf2000} 
M.J. Graf and A.V. Baratsky: 
Phys. Rev. B {\bf 62} (2000) 9697. 

\bibitem{ref:Wu2001}
W.C. Wu and R. Joynt:
Phys. Rev. B {\bf 64} (2001) 100507.

\bibitem{ref:Tanaka2003} Y. Tanaka, K. Kuroki, 
Y. Tanuma and S. Kashiwaya: 
J. Phys. Soc. Jpn. {\bf 72} (2003) 2157.
\label{ref:Tanaka2003}

\bibitem{ref:Kusunose2002} 
H. Kusunose and M. Sigrist: 
Europhys. Lett.{\bf 60} (2002) 281.
\label{ref:Kusunose2002}

\bibitem{ref:Udagawa2004}
M. Udagawa, Y. Yanase and M. Ogata: 
Phys. Rev. B {\bf 70} (2004) 184515. 

\bibitem{ref:Contreras2004} 
P. Contreras, M. Walker and K. Samokhin: 
Phys. Rev. B {\bf 70} (2004) 184528. 
\label{ref:Contreras2004}

\bibitem{ref:Hirschfeld1986} 
P.J. Hirschfeld, D. Vollhardt 
and P. W\"olfle: 
Solid State Commun. {\bf 59} (1986) 111. 

\bibitem{ref:Schmitt-Rink1986} 
S. Schmitt-Rink, K. Miyake and C.M. Varma: 
Phys. Rev. Lett. {\bf 57} (1986) 2575. 

\bibitem{ref:Hirschfeld1988} 
P.J. Hirschfeld, P. W\"olfle and D. Einzel: 
Phys. Rev. B {\bf 37} (1988) 83. 

\bibitem{ref:Walker2001} 
M.B. Walker, M.F. Smith and K.V. Samokhin: 
Phys. Rev. B {\bf 65} (2001) 014517. 
\label{ref:Walker2001}

\bibitem{ref:Oguchi1995} 
T. Oguchi: 
Phys. Rev. B {\bf 51} (1995) 1385. 

\bibitem{ref:Singh1995}
D.J. Singh: 
Phys. Rev. B {\bf 52} (1995) 1358. 

\bibitem{ref:Comment1}
This value of $t_5$ is a little smaller than 
that in ref.~\ref{ref:Nomura2002a}, 
for which the Fermi surfaces 
$\alpha$ and $\beta$ become more tetragonal 
than those in ref.~\ref{ref:Nomura2002a}. 
Calculated specific heat fits 
to the experimental data as well as 
in ref.~\ref{ref:Nomura2002a}. 

\bibitem{ref:Mackenzie1996} 
A.P. Mackenzie, S.R. Julien, A.J. Diver, G.J. McMullan, 
M.P. Ray, G.G. Lonzarich, Y. Maeno, S. Nishizaki and T. Fujita: 
Phys. Rev. Lett. {\bf 76} (1996) 3786. 

\bibitem{ref:Damascelli2000}
A. Damascelli, D.H. Lu, K.M. Shen, N.P. Armitage, F. Ronning, 
D.L. Feng, C. Kim, Z.X. Shen, T. Kimura, Y. Tokura, Z.Q. Mao 
and Y. Maeno: 
Phys. Rev. Lett. {\bf 85} (2000) 5194. 

\bibitem{ref:Rammer} For a textbook, 
J. Rammer: {\it Quantum Transport Theory} 
(Westview Press; Perseus Books Group, U.S., 2004). 

\bibitem{ref:Abrikosov} For a textbook, 
A.A. Abrikosov: {\it Fundamentals of the Theory of Metals} 
(Elsevier Science Publishers B.V., North-Holland, 1988).
\label{ref:Abrikosov} 

\bibitem{ref:Mineev} For a textbook, 
V.P. Mineev and K.V. Samokhin: 
{\it Introduction to Unconventional Superconductivity} 
(Gordon and Breach Science Publishers, New York, 1999). 

\bibitem{ref:Langer1962} 
J.S. Langer: 
Phys. Rev. {\bf 128} (1962) 110. 

\bibitem{ref:Luttinger1964} 
J.M. Luttinger: 
Phys. Rev. {\bf 135} (1964) A1505. 

\bibitem{ref:Mazin1999}
I.I. Mazin and D.J. Singh: 
Phys. Rev. Lett. {\bf 82} (1999) 4324. 

\bibitem{ref:Nomura2000a}
T. Nomura and K. Yamada: 
J. Phys. Soc. Jpn. {\bf 69} (2000) 1856. 
\label{ref:Nomura2000a}

\bibitem{ref:Eremin2002} 
I. Eremin, D. Manske and K.H. Bennemann: 
Phys. Rev. B {\bf 65} (2001) 220502. 

\bibitem{ref:Kikugawa2004} 
N. Kikugawa, C. Bergemann, A.P. Mackenzie and Y. Maeno: 
Phys. Rev. B {\bf 70} (2004) 134520. 

\bibitem{ref:Kuroki2001} 
K. Kuroki, M. Ogata, R. Arita and H. Aoki: 
Phys. Rev. B {\bf 63} (2001) 060506. 

\bibitem{ref:Kontani2004} 
H. Kontani: 
Phys. Rev. B {\bf 70} (2004) 054507. 

\bibitem{ref:Machida1999}
K. Machida, T. Nishira and T. Ohmi: 
J. Phys. Soc. Jpn. {\bf 68} (1999) 3364.

\bibitem{ref:Deguchi2004a} 
K. Deguchi, Z.Q. Mao, H. Yaguchi and Y. Maeno: 
Phys. Rev. Lett. {\bf 92} (2004) 047002. 

\bibitem{ref:Deguchi2004b} 
K. Deguchi, Z.Q. Mao and Y. Maeno: 
J. Phys. Soc. Jpn. {\bf 73} (2004) 1313.

\bibitem{ref:Kusunose2004} 
H. Kusunose: 
J. Phys. Soc. Jpn. {\bf 73} (2004) 2512.

\bibitem{ref:Mao2000} 
Z.Q. Mao, Y. Maeno, S. Nishizaki, T. Akima and T. Ishiguro: 
Phys. Rev. Lett. {\bf 84} (2000) 991. 

\bibitem{ref:Tanatar2001a} 
M.A. Tanatar, M. Suzuki, S. Nagai, Z.Q. Mao, 
Y. Maeno and T. Ishiguro: 
Phys. Rev. Lett. {\bf 86} (2001) 2649. 

\bibitem{ref:Yanase} 
Y. Yanase and M. Ogata: unpublished. 

\bibitem{ref:Yanase2003} 
Y. Yanase and M. Ogata: 
J. Phys. Soc. Jpn. {\bf 72} (2003) 673. 

\end{thebibliography}
\end{document}